\documentclass[11pt]{article}
\usepackage{amssymb,graphicx}
\usepackage{tabularx}
\usepackage{array}
\def\squarebox#1{\hbox to #1{\hfill\vbox to #1{\vfill}}}
\newcommand{\qed}{\hspace*{\fill}
            \vbox{\hrule\hbox{\vrule\squarebox{.667em}\vrule}\hrule}\smallskip\newline}
\newtheorem{THEOREM}{Theorem}
\newenvironment{theorem}{\begin{THEOREM} \hspace{-.85em} {\bf :} \rm}                        {\end{THEOREM}}
\newtheorem{LEMMA}{Lemma}
\newenvironment{lemma}{\begin{LEMMA} \hspace{-.85em} {\bf :} \rm}                      {\end{LEMMA}}
\newtheorem{COROLLARY}{Corollary}

\newenvironment{proof}{\noindent {\bf Proof:}}{\qed}
\newtheorem{DEFINITION}{Definition}

\newtheorem{CLAIM}{Claim}
                      {\end{CLAIM}}

\begin{document}

\title{Minkowski Sum Selection and Finding }

\author{Cheng-Wei Luo$^1$, Hsiao-Fei Liu$^1$, Peng-An Chen$^1$, and Kun-Mao Chao$^{1,2,3}$
\\\\
$^1$Department of Computer Science and Information Engineering \\
$^2$Graduate Institute of Biomedical Electronics and
Bioinformatics \\
$^3$Graduate Institute of Networking and Multimedia \\
National Taiwan University, Taipei, Taiwan 106\\
} \maketitle
\date

%
%

\begin{abstract}
Let $P,Q\subseteq {\mathbb{R}^{2}}$ be two $n$-point multisets and
$Ar\geq b$ be a set of $\lambda$ inequalities on $x$ and $y$, where
$A\in\mathbb{R}^{\lambda \times 2}$, $r=[^x_y]$, and $b\in
\mathbb{R}^{\lambda}$. Define the \textit{constrained Minkowski sum}
$(P\oplus Q)_{Ar\geq b}$ as the multiset $\{(p+q) | p \in P,q \in
Q,A(p+q)\geq b\}$. Given $P$, $Q$, $Ar\geq b$, an objective function
$f:\mathbb{R}^{2}\rightarrow\mathbb{R}$, and a positive integer $k$,
the \textsc{Minkowski Sum Selection} problem is to find the $k^{th}$
largest objective value among all objective values of points in
$(P\oplus Q)_{Ar\geq b}$. Given $P$, $Q$, $Ar\geq b$, an objective
function $f:\mathbb{R}^{2}\rightarrow\mathbb{R}$, and a real number
$\delta$, the \textsc{Minkowski Sum Finding} problem is to find a
point $(x^*,y^*)$ in $(P\oplus Q)_{Ar\geq b}$ such that
$|f(x^*,y^*)-\delta|$ is minimized. For the \textsc{Minkowski Sum
Selection} problem with linear objective functions, we obtain the
following results: (1) optimal $O(n\log n)$ time algorithms for
$\lambda=1$; (2) $O(n\log^2 n)$ time deterministic algorithms and
expected $O(n\log n)$ time randomized algorithms for any fixed
$\lambda>1$. For the \textsc{Minkowski Sum Finding} problem with
linear objective functions or objective functions of the form
 $f(x,y)=\frac{by}{ax}$, we construct optimal $O(n\log n)$ time algorithms for any fixed
$\lambda\geq 1$. As a byproduct, we obtain improved algorithms for
the \textsc{Length-Constrained Sum Selection} problem and the
\textsc{Density Finding} problem.
\end{abstract}
\vspace{-10pt} {\bf\quad\quad\  Keywords.} Bioinformatics, Sequence
analysis, Minkowski sum.
\section {Introduction}
Let $P,Q\subseteq {\mathbb{R}^{2}}$ be two $n$-point multisets and
$Ar\geq b$ be a set of $\lambda$ inequalities on $x$ and $y$, where $A\in\mathbb{R}^{\lambda \times 2}$, $r=[^x_y]$, and $b\in \mathbb{R}^{\lambda}$. Define the \textit{constrained Minkowski
sum} $(P\oplus Q)_{Ar\geq b}$ as the multiset $\{(p+q) | p \in P,q \in Q,A(p+q)\geq b\}.$

In the \textsc{Minkowski Sum Optimization} problem, we are given
$P$, $Q$, $Ar\geq b$, and an objective function
$f:\mathbb{R}^{2}\rightarrow\mathbb{R}$. The goal is to find the
maximum objective value among all objective values of points in
$(P\oplus Q)_{Ar\geq b}$. A function $f:D\subseteq
\mathbb{R}^2\rightarrow\mathbb{R}$ is said to be
\textit{quasiconvex} if and only if for all points $v_1,v_2 \in D$
and all $\gamma \in [0,1]$, one has $f(\gamma \cdot v_1 + (1 -
\gamma) \cdot v_2) \leq \max(f(v_1), f(v_2))$. Bernholt
\textit{et~al.}~\cite{Bern} studied the \textsc{Minkowski Sum
Optimization} problem for quasiconvex objective functions and showed
that their results have applications to many optimization problems
arising in computational
biology~\cite{All,starmap,Gold,Huang,LJC,LHJC,Lip,Wang}. In this
paper, two variations of the \textsc{Minkowski Sum Optimization}
problem are studied: the \textsc{Minkowski Sum Selection} problem
and the \textsc{Minkowski Sum Finding} problem.

In the \textsc{Minkowski Sum Selection} problem, we are given $P$,
$Q$, $Ar\geq b$, an objective function
$f:\mathbb{R}^{2}\rightarrow\mathbb{R}$, and a positive integer $k$.
The goal is to find the $k^{th}$ largest objective value among all
objective values of points in $(P\oplus Q)_{Ar\geq b}$. The
\textsc{Minkowski Sum Optimization} problem is equivalent to the
\textsc{Minkowski Sum Selection} problem with $k=1$. A variety of
selection problems, including the \textsc{Sum Selection}
problem~\cite{Bengtsson,TC06}, the \textsc{Length-Constrained Sum
Selection} problem~\cite{TC05}, and the \textsc{Slope Selection}
problem~\cite{Cole,Ma}, are linear-time reducible to the
\textsc{Minkowski Sum Selection} problem with a linear objective
function or an objective function of the form $f(x,y) =
\frac{by}{ax}$. It is desirable that relevant selection problems
from diverse fields are integrated into a single one, so we don't
have to consider them separately. Next, let us look at the use of
the \textsc{Minkowski Sum Selection} problem in practice. As
mentioned above, the \textsc{Minkowski Sum Optimization} problem
finds applications to many optimization problems arising in
computational
biology~\cite{All,starmap,Gold,Huang,LJC,LHJC,Lip,Wang}. In these
optimization problems, the objective functions are chosen such that
feasible solutions with higher objective values are ``more likely''
to be biologically meaningful. However, it is not guaranteed that
the best feasible solution always satisfies the needs of biologists.
If the best feasible solution does not interest biologists or does
not provide enough information, we still have to find the second
best feasible solution, the third best feasible solution and so on
until a satisfying feasible solution is finally found. As a result,
it is desirable to know how to dig out extra good feasible solutions
in case that the best feasible solution is not sufficient.

In the \textsc{Minkowski Sum Finding} problem, we are given $P$,
$Q$, $Ar\geq b$, an objective function
$f:\mathbb{R}^{2}\rightarrow\mathbb{R}$, and a real number $\delta$.
The goal is to find a point $(x^*,y^*)$ in $(P\oplus Q)_{Ar\geq b}$
such that $|f(x^*,y^*)-\delta|$ is minimized. This problem
originates from the study of the \textsc{Density Finding} problem
proposed by Lee~$et~al.$~\cite{LLL}. The \textsc{Density Finding}
problem can be regarded as a specialization of the \textsc{Minkowski
Sum Finding} problem with objective function $f(x,y)=\frac{y}{x}$
and find applications in recognizing promoters in DNA
sequences~\cite{Ioshikhes,Ohler}. In these applications, the goal is
not to find the feasible solution with the highest objective value.
Instead, feasible solutions with objective values close to some
specific number, say $\delta$, are thought to be more biologically
meaningful and preferred.

The main results obtained in this paper are as follows.
\vspace{-3pt}

\begin{itemize}
        \item The \textsc{Minkowski Sum Selection} problem
        with one constraint and a linear objective function can be solved in optimal $O(n\log n)$ time.
        \item The \textsc{Minkowski Sum Selection} problem
        with two constraints and a linear objective function can be solved in $O(n\log^2 n)$
        time by a deterministic algorithm and expected $O(n\log n)$ time by a randomized algorithm.
        \item For any fixed $\lambda>2$, the \textsc{Minkowski Sum Selection} problem
        with $\lambda$ constraints and a linear objective
        function is shown to be asymptotically equivalent to the \textsc{Minkowski Sum Selection} problem
        with two constraints and a linear objective function.
        \item The \textsc{Minkowski Sum Finding} problem with any fixed number of constraints can be solved in
        optimal $O(n\log n)$ time if the objective function $f(x,y)$ is linear or of the form $\frac{by}{ax}$.
\end{itemize}
\vspace{-5pt}

As a byproduct, we obtain improved algorithms for the
\textsc{Length-Constrained Sum Selection} problem~\cite{TC05} and
the \textsc{Density Finding} problem~\cite{LLL}. Recently, Lin and
Lee~\cite{TC05} proposed an expected $O(n\log(u-l+1))$-time
randomized algorithm for the \textsc{Length-Constrained Sum
Selection} problem, where $n$ is the size of the input instance and
$l,u\in \mathbb{N}$ are two given parameters with $1\leq l< u\leq
n$. In this paper, we obtain a worst-case $O(n\log(u-l+1))$-time
deterministic algorithm for the \textsc{Length-Constrained Sum
Selection} problem (see Appendix A). Lee, Lin, and Lu~\cite{LLL} showed the \textsc{Density Finding} problem has a lower bound of $\Omega(n\log n)$ and proposed an
$O(n\log^2 m)$-time algorithm for it, where $n$ is the size of the
input instance and $m$ is a parameter whose value may be as large
as~$n$. In this paper, we give an optimal $O(n\log n)$-time algorithm for
the \textsc{Density Finding} problem (see Appendix B).

\section{Preliminaries}
In the following, we review some definitions and theorems. For more
details, readers can refer to \cite{Bern,Frede}. A matrix $X\in
\mathbb{R}^{\mathfrak{n} \times \mathfrak{m}}$ is said to be
\textit{sorted} if the values of each row and each column are in
nondecreasing order. Frederickson and Johnson \cite{Frede} gave some
results about the selection problem and the ranking problem in a
collection of sorted matrices. From the results of Frederickson and
Johnson~\cite{Frede}, we have the following theorems.

\begin{theorem}\label{SelectionTime}
The selection problem in a collection of sorted matrices is given a
rank $k$ and a collection of sorted matrices $\{X_1,\ldots,X_N\}$ in
which $X_j$ has dimensions $n_j\times m_j$, $n_j\geq m_j$, to find
the $k^{th}$ largest element among all elements of sorted matrices
in $\{X_1,\ldots,X_N\}$. This problem is able to be solved in
$O(\sum_{j=1}^Nm_j\log(2n_j/m_j))$ time.
\end{theorem}

\begin{theorem}\label{RankingTime}
The ranking problem in a collection of sorted matrices is given an
element and a collection of sorted matrices $\{X_1,\ldots,X_N\}$ in
which $X_j$ has dimensions $n_j\times m_j$, $n_j\geq m_j$, to find
the rank of the given element among all elements of sorted matrices
in $\{X_1,\ldots,X_N\}$. This problem can be solved in
$O(\sum_{j=1}^Nm_j\log(2n_j/m_j))$ time.
\end{theorem}

By the recent works of Bernholt~\textit{et al.}~\cite{Bern}, we have
the following theorems.
\begin{theorem}\label{Convexhull}
Given a set of $\lambda$ linear inequalities $Ar \geq b$ and two
$n$-point multisets $P$, $Q \subseteq{\mathbb{R}^{2}}$, one can
compute the vertices of the convex hull of $(P\oplus Q)_{Ar\geq b}$
in $O(\lambda \log \lambda + \lambda \cdot n \log n)$ time.
\end{theorem}

\begin{theorem}\label{ConvexhullLB}
The problem of maximizing a quasiconvex objective function $f$ over
the constrained Minkowski sum $(P\oplus Q)_{Ar\geq b}$ requires
$\Omega(n \log n)$ time in the algebraic decision tree model even if
$f$ is a linear function and $Ar \geq b$ consists of only one
constraint.
\end{theorem}


\section{Minkowski Sum Selection with One Constraint}
In this section we study the \textsc{Minkowski Sum Selection}
problem and give an optimal $O(n\log n)$ time algorithm for the case
where only one linear constraint is given and the objective function
is also linear.

\subsection{Input Transformation}
Given $P=\{(x_{1,1},y_{1,1}),\ldots,(x_{1,n},y_{1,n})\}$,
$Q=\{(x_{2,1},y_{2,1}),\ldots,(x_{2,n},y_{2,n})\}$, a positive
integer $k$, one constraint $L$: $ax+by\geq c$, and a linear
objective function $f(x,y)=dx+ey$, where $a$, $b$, $c$, $d$, and $e$
are all constants, we perform the following transformation.
\vspace{-5pt}

\begin{enumerate}
    \item Change the content of $P$ and $Q$ to
    $\{(ax_{1,1}+by_{1,1},dx_{1,1}+ey_{1,1}),\ldots,(ax_{1,n}+by_{1,n},dx_{1,n}+ey_{1,n})\}$,
    and
    $\{(ax_{2,1}+by_{2,1},dx_{2,1}+ey_{2,1}),\ldots,(ax_{2,n}+by_{2,n},dx_{2,n}+ey_{2,n})\}$, respectively.
    \item Change the constraint from $ax+by \geq c$ to $x\geq c$.
    \item Change the objective function from $dx+ey$ to $y$.
\end{enumerate}
\vspace{-5pt}

This transformation can be done in $O(n)$ time and the answer
remains the same. Hence from now on, our goal becomes to find the
$k^{th}$ largest $y$-coordinate on the constrained Minkowski sum of
$P$ and $Q$ subject to the constraint $L:x\geq c$.

\subsection{Algorithm}
For ease of exposition, we assume that no two points in $P$ and $Q$
have the same $x$-coordinate and $n$ is a power of two. The
algorithm proceeds as follows. First, we sort $P$ and $Q$ into $P_x$
and $Q_x$ ($P_y$ and $Q_y$, respectively) in nondecreasing order of
$x$-coordinates ($y$-coordinates, respectively) in $O(n \log n)$
time. Next, we use a divide-and-conquer approach to store the
$y$-coordinates of $(P\oplus Q)_{x\geq c}$ as a collection of sorted
matrices and then apply Theorem~\ref{SelectionTime} to select the
$k^{th}$ largest element from the elements of these sorted matrices.

Now we explain how to store the $y$-coordinates of $(P\oplus
Q)_{x\geq c}$ as a collection of sorted matrices. Let $P_x =
((x_1,y_1),\ldots,(x_n,y_n))$, $Q_x =
((\bar{x}_1,\bar{y_1}),\ldots,(\bar{x}_n,\bar{y}_n))$, $P_y =
((x'_1,y'_1),\ldots,(x'_n,y'_n)),$ and $Q_y =
((\bar{x}'_1,\bar{y}'_1),\ldots,(\bar{x}'_n,\bar{y}'_n)).$ We then
divide $P_x$ into two halves of equal size: $A =
((x_1,y_1),\ldots,(x_{n/2},y_{n/2}))$ and $B=
((x_{n/2+1},y_{n/2+1}),\ldots,(x_n,y_n)).$ Find a point
$(\bar{x}_t,\bar{y}_t)$ of $Q_x$ such that $x_{n/2}+\bar{x}_{t}<c$
and $t$ is maximized. Then divide $Q_x$ into two halves:
$C=((\bar{x}_1,\bar{y}_1),\ldots,(\bar{x}_{t},\bar{y}_{t}))$ and
$D=((\bar{x}_{t+1},\bar{y}_{t+1}),\ldots,(\bar{x}_{n},\bar{y}_{n})).$
The set $(P\oplus Q)_{x\geq c}$ is the union of $(A\oplus C)_{x\geq
c}$, $(A\oplus D)_{x\geq c}$, $(B\oplus C)_{x\geq c}$, and $(B\oplus
D)_{x \geq c}$. Because $\bar{x}_t$ is the largest $x$-coordinate
among all $x$-coordinates of points in $Q_x$ such that
$x_{n/2}+\bar{x}_{t}<c$, we know that all points in $A\oplus C$
cannot satisfy the constraint $x\geq c$. Hence, we only need to
consider points in $A\oplus D$, $B\oplus C$, and $B\oplus D$.
Because $P_x$ and $Q_x$ are in nondecreasing order of
$x$-coordinates, it is guaranteed that all points in $B\oplus D$
satisfy the constraint $L$, i.e., $B\oplus D = (B\oplus D)_{x\geq
c}$. Construct in linear time \textit{row\_vector} =
$(r_1,r_2,\ldots,r_{n/2})$ which is the $y$-coordinates in the
subsequence of $P_y$ resulting from removing points with
$x$-coordinates no greater than $x_{n/2}$ from $P_y$. Construct in
linear time \textit{column\_vector} = $(c_1,c_2,\ldots,c_{n-t})$
which is the $y$-coordinates in the subsequence of $Q_y$ resulting
from removing points with $x$-coordinates no greater than
$\bar{x}_{t}$ from $Q_y$. Note \textit{row\_vector} is the same as
the result of sorting $B$ into nondecreasing order of
$y$-coordinates, and \textit{column\_vector} is the same as the
result of sorting $D$ into nondecreasing order of $y$-coordinates.
Thus, we have $\{y: (x,y)\in (B\oplus D)_{x\geq c}\}=\{y: (x,y)\in
B\oplus D\} = \{r_i+c_j: 1\leq i \leq |row\_vector|,1\leq j \leq
|column\_vector|\}$. Therefore, we can store the $y$-coordinates of
$B\oplus D = (B\oplus D)_{x\geq c}$ as a sorted matrix $X$ of
dimensions $|row\_vector|\times|column\_vector|$ where the
$(i,j)$-th element of $X$ is $r_i +c_j$. Note that it is not
necessary to explicitly construct the sorted matrix $X$, which needs
$\Omega(n^2)$ time. Because the $(i,j)$-th element of $X$ can be
obtained by summing up $r_i$ and $c_j$, we only need to keep
$row\_vector$ and $column\_vector$. The rest is to construct the
sorted matrices for the $y$-coordinates of points in $(A\oplus
D)_{x\geq c}$ and $(B\oplus C)_{x\geq c}$. It is accomplished by
applying the above approach recursively. The pseudocode is shown in
Figure~\ref{algo1} and Figure~\ref{algo2}. We now analyze the time
complexity.

\begin{figure}[!h]
    \noindent\hrulefill\vspace{-0.6cm}
    \begin{tabbing}
    \hspace*{1em} \= \hspace*{1em} \= \hspace*{1em} \= \hspace*{1em}
    \=
    \hspace*{1em} \kill \\
    \textbf{Algorithm} {ConstructMatrices$(P_x,Q_x,P_y,Q_y,L$: $x\geq c)$}\\
    \textbf{Input:} $P_x$ and $P_y$ are the results of sorting the multiset $P\subseteq \mathbb{R}^2$ in nondecreasing\\
    \>order of $x$-coordinates and $y$-coordinates, respectively. $Q_x$ and $Q_y$ are the results\\
    \>of sorting the multiset $Q\subseteq \mathbb{R}^2$ in nondecreasing order of $x$-coordinates and\\
    \>$y$-coordinates, respectively. A linear constraint $L$: $x\geq c$.\\
    \textbf{Output:} The $y$-coordinates of points in $(P\oplus Q)_{x\geq c}$ as a collection of sorted matrices.\\

    \ 1\> $n' \leftarrow |P_x|$; $m' \leftarrow |Q_x|$.\\
    \ 2\> Let $P_x = ((x_1,y_1),\ldots,(x_{n'},y_{n'}))$,
    $Q_x
    =((\bar{x}_1,\bar{y_1}),\ldots,(\bar{x}_{m'},\bar{y}_{m'}))$,\\
    \>$P_y = ((x'_1,y'_1),\ldots,(x'_{n'},y'_{n'})),$ and
    $Q_y =
    ((\bar{x}'_1,\bar{y}'_1),\ldots,(\bar{x}'_{m'},\bar{y}'_{m'})).$\\
    \ 3\> $\bar{x}_{0}\leftarrow-\infty$.\\
    \ 4\> \textbf{if} $n'\leq 0$ or $m'\leq 0$ \textbf{then}\\
    \ 5\qquad \textbf{return}\\
    \ 6\> \textbf{if} $n'=1$ or $m'=1$ \textbf{then}\\
    \ 7\qquad Scan points in $P_x\oplus Q_x$ to find all points satisfying $L$ and construct the sorted\\
    \quad\qquad matrix for $y$-coordinates of these points.\\
    \ 8\qquad \textbf{return} the above sorted matrix.\\
    \ 9\> \textbf{for} $t\leftarrow m'$ down to $0$ \textbf{do}\\
     10\qquad \textbf{if} $x_{n'/2}+\bar{x}_t<c$ \textbf{then}\\
     11\>\>\> \textit{row\_vector} $\leftarrow$ subsequence of $P_y$ being removed points with\\
     \>\>\> $x$-coordinates $\leq x_{n'/2}$.\\
     12\>\>\> \textit{column\_vector} $\leftarrow$ subsequence of $Q_y$ being removed points with\\
     \>\>\> $x$-coordinates $\leq \bar{x}_{t}$.\\
     13\>\>\> $A_x \leftarrow P_x[1,n'/2]$; $B_x \leftarrow P_x[n'/2+1,n']$; $C_x \leftarrow Q_x[1,t]$; $D_x \leftarrow Q_x[t+1,m']$.\\
     14\>\>\> $A_y \leftarrow$ subsequence of $P_y$ being removed points with $x$-coordinates $>x_{n'/2}$.\\
     15\>\>\> $B_y \leftarrow$ subsequence of $P_y$ being removed points with $x$-coordinates $\leq x_{n'/2}$.\\
     16\>\>\> $C_y \leftarrow$ subsequence of $Q_y$ being removed points with $x$-coordinates $>\bar{x}_{t}$.\\
     17\>\>\> $D_y \leftarrow$ subsequence of $Q_y$ being removed points with $x$-coordinates $\leq \bar{x}_{t}$.\\
     18\>\>\> ConstructMatrices($A_x,D_x,A_y,D_y,L$: $x\geq c$).\\
     19\>\>\> ConstructMatrices($B_x,C_x,B_y,C_y,L$: $x\geq c$).\\
     20\>\>\> \textbf{return} \textit{row\_vector} and \textit{column\_vector}.\\
    \end{tabbing}
    \vspace{-0.75 cm}\noindent\hrulefill
    \caption{The subroutine for the \textsc{Minkowski Sum Selection} problem with one linear constraint and a linear objective function.}
    \label{algo1}
    \end{figure}

\begin{figure}[!h]
    \noindent\hrulefill\vspace{-0.6cm}
    \begin{tabbing}
    \hspace*{1em} \= \hspace*{1em} \= \hspace*{1em} \= \hspace*{1em}
    \=
    \hspace*{1em} \kill \\
    \textbf{Algorithm} {Selection$_1(P,Q,L,f,k)$}\\
    \textbf{Input:} Two multisets $P\subseteq \mathbb{R}^2$ and $Q\subseteq \mathbb{R}^2$; a linear constraint
    $L$; a linear objective\\
    \>function  $f:{\mathbb{R}^{2}}\rightarrow \mathbb{R}$; a positive integer $k$.\\
    \textbf{Output:} The $k^{th}$ largest value among
    all objective values of points in $(P\oplus Q)_{L}$.\\

    \ 1\> Perform the input transformation described in Section~3.1.\\
    \ 2\> Sort $P$ and $Q$ into $P_x$ and $Q_x$, respectively, in nondecreasing order of $x$-coordinates.\\
    \ 3\> Sort $P$ and $Q$ into $P_y$ and $Q_y$, respectively, in nondecreasing order of $y$-coordinates.\\
    \ 4\> $S \leftarrow$ ConstructMatrices$(P_x,Q_x,P_y,Q_y,L)$.\\
    \ 5\> \textbf{return} the $k^{th}$ largest element among the elements of sorted matrices in $S.$\\
    \end{tabbing}
    \vspace{-0.75 cm}\noindent\hrulefill
    \caption{The main procedure for the \textsc{Minkowski Sum Selection} problem with one linear constraint and a linear objective function.}
    \label{algo2}
    \end{figure}

\begin{lemma}\label{IndexSum}
Given a matrix $X\in \mathbb{R}^{\mathfrak{N}\times \mathfrak{M}}$,
we define the \textit{side length} of $X$ be $\mathfrak{N}+
\mathfrak{M}$. Letting $T(n',m')$ be the running time of
ConstructMatrices($P_x,Q_x,P_y,Q_y,L$), where $n'=|P_x|=|P_y|$ and
$m'=|Q_x|=|Q_y|$, we have $T(n',m')= O((n'+m')\log (n'+1)).$
Similarly, letting $M(n',m')$ be the sum of the side lengths of all
sorted matrices created by running
ConstructMatrices($P_x,Q_x,P_y,Q_y,L$), we have $M(n',m')=
O((n'+m')\log (n'+1)).$
\end{lemma}
\begin{proof}
It suffices to prove that $T(n',m')= O((n'+m')\log (n'+1)).$ By
Algorithm ConstructMatrices in Figure~\ref{algo1}, we have
$T(n',m')\leq \max_{0\leq i\leq
m'}\{c'(n'+m')+T(n'/2,i)+T(n'/2,m'-i)\} \mbox{ for some constant }
c'.$
Then by induction on $n'$, it is easy to prove that $T(n',m')$ is
$O((n'+m')\log (n'+1))$.
\end{proof}

\begin{theorem}\label{OneConstraint}
Given two $n$-point multisets $P\subseteq \mathbb{R}^2$ and
$Q\subseteq \mathbb{R}^2$, a positive integer $k$, a linear
constraint $L$, and a linear objective function
$f:\mathbb{R}^{2}\rightarrow\mathbb{R}$, Algorithm Selection$_1$
finds the $k^{th}$ largest objective value among all objective
values of points in $(P\oplus Q)_L$ in $O(n\log n)$ time. Hence, by
Theorem~\ref{ConvexhullLB}, Algorithm Selection$_1$ is optimal.
\end{theorem}
\begin{proof}
Let $S=\{X_1,\ldots,X_N\}$ be the sorted matrices produced at Step~4
in Algorithm Selection$_1$. Let $X_j$, $1\leq j \leq N$, be of
dimensions $n_j\times m_j$ where $n_j\geq m_j$. By
Lemma~\ref{IndexSum}, we have $O(\sum_{i=1}^{N}(m_i+n_i))=O(n\log
n).$ By Theorem~\ref{SelectionTime}, the time required to find the
$k^{th}$ largest element from the elements of matrices in $S$ is
$O(\sum_{i=1}^{N}m_i\log(2n_i/m_i))$. Since
\begin{displaymath}
\sum_{i=1}^{N}m_i\log(2n_i/m_i)\leq
\sum_{i=1}^{N}m_i\frac{2n_i}{m_i}=\sum_{i=1}^{N}2n_i\leq
\sum_{i=1}^{N}2(m_i+n_i) = O(n\log n),
\end{displaymath}
the time for selecting the $k^{th}$ largest element from elements of
matrices in $S$ is $O(n\log n)$. Combining this with the time for
the input transformation, sorting, and executing
ConstructMatrices($P_x,Q_x,P_y,Q_y,L$), we conclude that the total
running time is $O(n\log n)$.
\end{proof}

Using similar techniques, the following problem can also be solved
in $O(n\log n)$ time. Given two $n$-point multisets $P\subseteq
\mathbb{R}^2$ and $Q\subseteq \mathbb{R}^2$, a linear constraint
$L$, a linear objective function
$f:\mathbb{R}^{2}\rightarrow\mathbb{R}$, and a real number $t$, the
problem is to find the rank of $t$ among all objective values of
points in $(P\oplus Q)_L$, where the rank of $t$ is equal to the
number of elements in $\{y|(x,y)\in (P\oplus Q)_L,y>t\}$ plus one.
The pseudocode is given in Figure~\ref{algo3}. Note that in
Algorithm Selection$_1$ and Algorithm Ranking$_1$, we assume the
input constraint is of the form $ax+by\geq c$. After slight
modifications, we can also cope with constraints of the form
$ax+by>c$. To avoid redundancy, we omit the details here. For ease
of exposition, we assume that Algorithm Selection$_1$ and Algorithm
Ranking$_1$ are also capable of coping with constraints of the form
$ax+by>c$ in the following sections.

\begin{figure}[!h]
    \noindent\hrulefill\vspace{-0.6cm}
    \begin{tabbing}
    \hspace*{1em} \= \hspace*{1em} \= \hspace*{1em} \= \hspace*{1em}
    \=
    \hspace*{1em} \kill \\
    \textbf{Algorithm} {Ranking$_1(P,Q,L,f,t)$}\\
    \textbf{Input:} Two multisets $P\subseteq \mathbb{R}^2$ and $Q\subseteq \mathbb{R}^2$; a linear constraint $L$; a linear objective\\
    \>function $f:{\mathbb{R}^{2}}\rightarrow \mathbb{R}$; a real number $t$.\\
    \textbf{Output:} The rank of $t$ among the objective values of points in $(P\oplus
    Q)_L$.\\

    \ 1\> Perform the input transformation in Section~3.1.\\
    \ 2\> Sort $P$ and $Q$ into $P_x$ and $Q_x$, respectively, in nondecreasing order of $x$-coordinates.\\
    \ 3\> Sort $P$ and $Q$ into $P_y$ and $Q_y$, respectively, in nondecreasing order of $y$-coordinates.\\
    \ 4\> $S \leftarrow$ ConstructMatrices$(P_x,Q_x,P_y,Q_y,L)$.\\
    \ 5\> \textbf{return} the rank of $t$ among the elements of sorted matrices in $S.$\\
    \end{tabbing}
    \vspace{-0.75 cm}\noindent\hrulefill
    \caption{The ranking algorithm for the Minkowski sum with one linear constraint and a linear objective function.}
    \label{algo3}
    \end{figure}


\section{Minkowski Sum Selection with Two Constraints}

In this section, we show the \textsc{Minkowski Sum Selection}
problem can be solved in worst-case $O(n\log^2 n)$ time and expected
$O(n\log n)$ time for the case where two linear constraints are
given and the objective function is linear.

\subsection{Input Transformation.}
\label{sec:transformation2}
Given
$P=\{(x_{1,1},y_{1,1}),\ldots,(x_{1,n},y_{1,n})\}$,
$Q=\{(x_{2,1},y_{2,1}),\ldots,(x_{2,n},y_{2,n})\}$, a positive
integer $k$, two constraints $L_1$: $a_1x+b_1y\geq c_1$ and $L_2$:
$a_2x+b_2y\geq c_2$, and a linear objective function $f(x,y)=dx+ey$,
where $a_1$, $b_1$, $c_1$, $a_2$, $b_2$, $c_2$, $d$, and $e$ are all
constants, we perform the following transformation.
\vspace{-5pt}

\begin{enumerate}
    \item Change the content of $P$ and $Q$ to
    $\{(a_1x_{1,1}+b_1y_{1,1},dx_{1,1}+ey_{1,1}),\ldots,(a_1x_{1,n}+b_1y_{1,n},dx_{1,n}+ey_{1,n})\}$,
    and
    $\{(a_1x_{2,1}+b_1y_{2,1},dx_{2,1}+ey_{2,1}),\ldots,(a_1x_{2,n}+b_1y_{2,n},dx_{2,n}+ey_{2,n})\}$, respectively.
    \item Change the constraints from $a_1x+b_1y \geq c_1$ and $a_2x+b_2y\geq c_2$ to $x\geq c_1$ and
            $\frac{a_2e-b_2d}{a_1e-b_1d}x+\frac{a_1b_2-b_1a_2}{a_1e-b_1d}y\geq c_2$,
            respectively.
    \item Change the objective function from $dx+ey$ to $y$.
\end{enumerate}
\vspace{-5pt}

This transformation can be done in $O(n)$ time and the answer
remains the same. Hence from now on, our goal becomes to find the
$k^{th}$ largest $y$-coordinate on the constrained Minkowski sum of
$P$ and $Q$ subject to the constraints $L_1$: $x\geq c_1$ and $L_2$:
$ax+by\geq c_2$, where $a=\frac{a_2e-b_2d}{a_1e-b_1d}$ and
$b=\frac{a_1b_2-b_1a_2}{a_1e-b_1d}.$ Note that if the two
constraints and the objective function are parallel, we cannot use
the above transformation. However, if the two constraints are
parallel, this problem can be solved in $O(n \log n)$ time. For the
space limitation, we present the algorithm for this special case in
Appendix~C.

\subsection{Algorithm}
After applying the above input transformation to our problem
instances, there are four possible cases: (1) $a<0,b<0$; (2)
$a>0,b>0$; (3) $a<0,b>0$; (4) $a>0,b<0$. Note that the two
constraints are not parallel implies $b \neq 0$. If $a=0$, we can
solve this case more easily in $O(n\log^2 n)$ time by using the same
technique stated later and we omit the details here. In the
following discussion we focus on Case (1), and the other three cases
can be solved in a similar way.

\begin{figure}[ht]
\begin{minipage}[t]{10cm}
\includegraphics[scale=0.5]{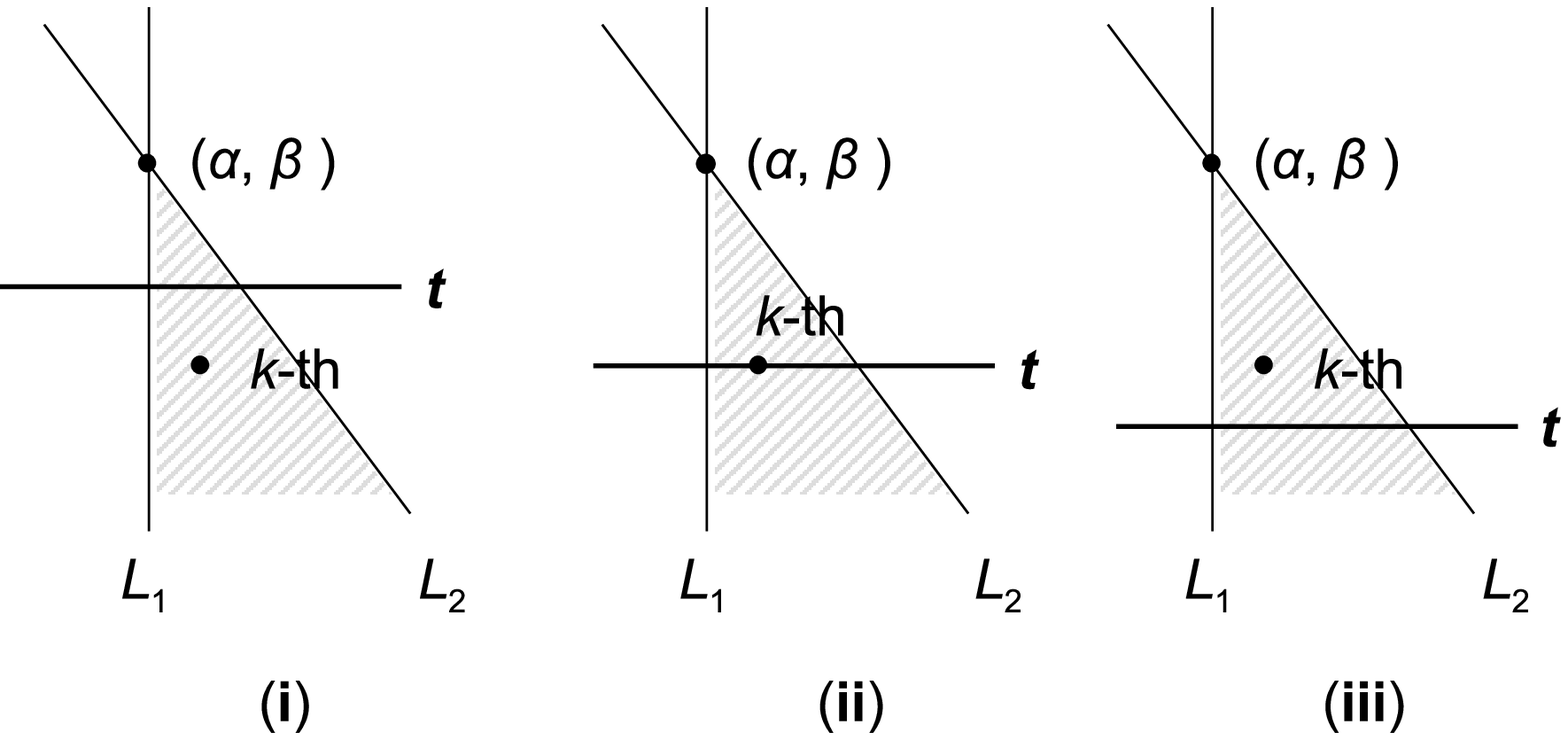}
\caption{The three possible cases for a given value $t$.}
\label{binarysearch}
\end{minipage}
\hspace{0.5cm}
\begin{minipage}[t]{5.5cm}
\includegraphics[scale=0.35]{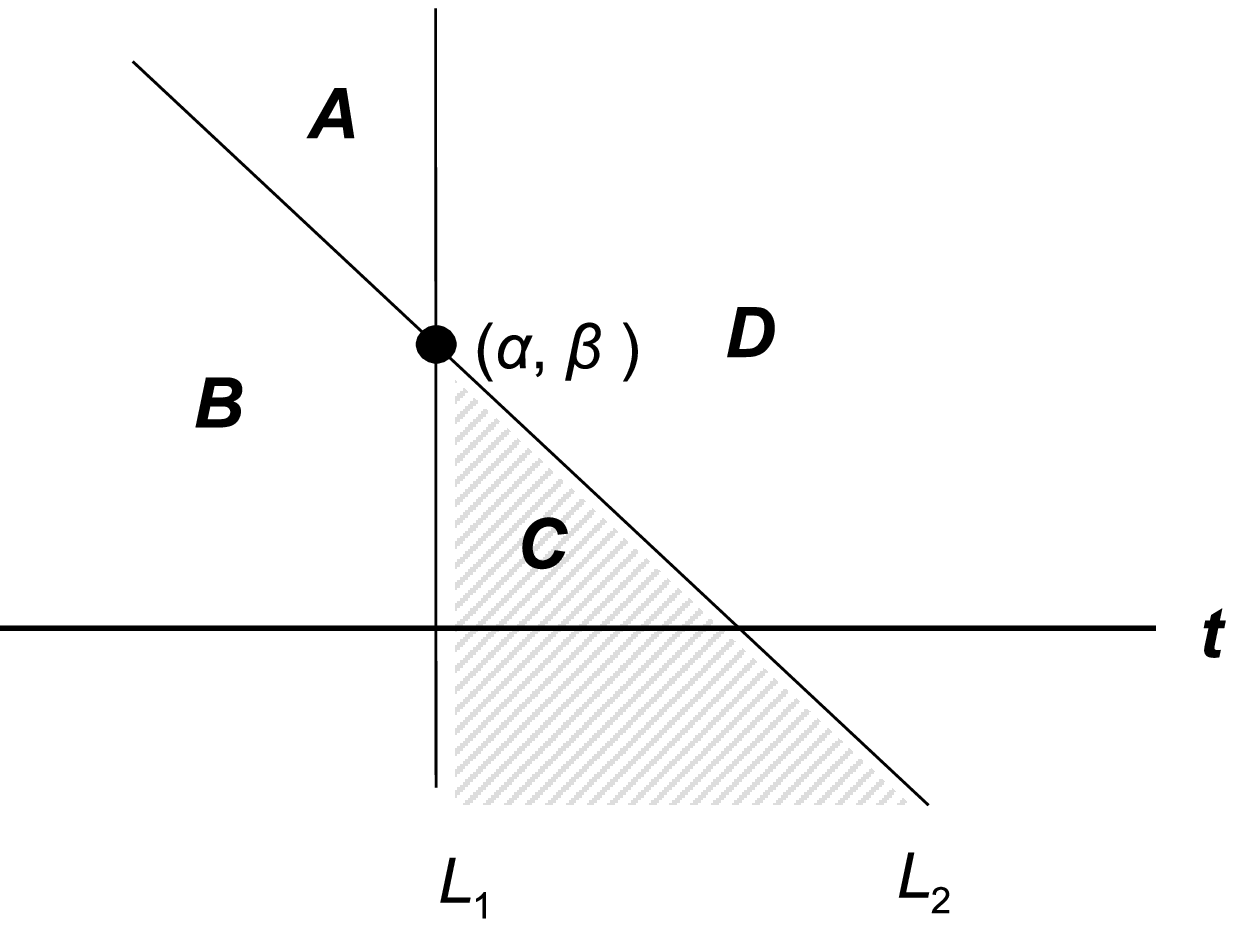}
\caption{The four regions above (exclusive) the line $y=t.$}
\label{calculateranking}
\end{minipage}
\end{figure}

For simplicity, we assume that $n$ is a power of two, and each point
in $(P\oplus Q)_{L_1,L_2}$ has a distinct $y$-coordinate. Now we are
ready to describe our algorithm. First, we sort $P$ and $Q$ into
$P_y$ and $Q_y$, respectively, in nondecreasing order of
$y$-coordinates using $O(n\log n)$ time. Let
$P_y=\{(x'_1,y'_1),\ldots,(x'_n,y'_n)\}$ and
$Q_y=\{(\bar{x}'_1,\bar{y}'_1),\ldots,(\bar{x}'_n,\bar{y}'_n)\}$.
Denote by $Y$ the sorted matrix of dimensions $n\times n$ where the
$(i,j)$-th element is $y'_i+\bar{y}'_j$. We then run a loop where an
integer interval $[l,u]$ is maintained such that the solution is
within the set $\{\mbox{the }h^{th} \mbox{ largest element of }Y:
l\leq h\leq u\}$. Initially, we set $l=1$ and $u=n^2$. At the
beginning of each iteration, we select the $\frac{u-l+1}{2}$-th
largest element $t$ of $Y$, which can be done in $O(n)$ time by
Theorem~\ref{SelectionTime}. Let $R$ be the rank of $t$ among the
objective values of points in $(P\oplus Q)_{L_1,L_2}$. Then there
are three possible cases: (i) $R< k$; (ii) $R= k$; (iii) $R > k$.
See Figure~\ref{binarysearch} for an illustration. If it is
Case~(i), then we reset $l$ to $\frac{u-l+1}{2}$ and continue the
next iteration. If it is Case~(ii), then we apply the algorithm for
the \textsc{Minkowski Sum Finding} problem (discussed in Section~6)
to find the point $p=(x^*,y^*)$ in $(P\oplus Q)_{L_1,L_2,y\leq t}$
in $O(n\log n)$ time such that $y^*$ is closest to $t$ and return
$y^*$. If it is Case~(iii), then we reset $u$ to $\frac{u-l+1}{2}$
and continue the next iteration.

It remains to describe the subroutine for computing $R$. Let $A =
\{(x,y)| x<c_1\mbox{ and }ax+by<c_2\}$, $B = \{(x,y)| x<c_1 \mbox{
and }ax+by \geq c_2 \mbox{ and } y>t\}$, $C = \{(x,y)| x\geq c_1
\mbox{ and }ax+by \geq c_2 \mbox{ and } y>t\}$ and $D = \{(x,y)|
x\geq c_1 \mbox{ and }ax+by<c_2 \mbox{ and } y>t\}$. See
Figure~\ref{calculateranking} for an illustration. First, we compute
the number of points in $(P\oplus Q)\cap(A\cup B)$, say $R_1$, by
calling Ranking$_1(P,Q,L'$: $x< c_1,f'(x,y)=y,t)-1$. Secondly, we
compute the number of points in $(P\oplus Q)\cap(A\cup D)$, say
$R_2$, by calling Ranking$_1(P,Q,L''$: $ax+by< c_2,f'(x,y)=y,t)-1.$
Thirdly, we compute the number of points in $(P\oplus Q)\cap A$, say
$R_3$, by calling Ranking$_1(P,Q,L''$: $ax+by<
c_2,f''(x,y)=-x,c_1)-1$. Finally, we compute the number of points in
$(P\oplus Q)_{y>t}$, say $R_t$. It can be done by applying
Theorem~\ref{RankingTime} to calculate the rank of $t$ among the
values of the elements in $Y$, say $R_t'$, and set $R_t$ to
$R_t'-1$. After getting $R_1$, $R_2$, $R_3$, and $R_t$, we can
compute $R$ by the following equation: $R=R_t-R_1-R_2+R_3+1$. Since
all $R_1,R_2,R_3,$ and $R_t$ can be computed in $O(n\log n)$ time,
the time for computing $R$ is $O(n\log n)$.

Now let us look at the total time complexity. Since the loop
consists of at most $O(\log n)$ iterations and each iteration takes
$O(n\log n)$ time, the total time complexity of the loop is
$O(n\log^2n)$. By combining this with the time for the input
transformation and sorting, we have the following theorem.

\begin{theorem}\label{Selection2time}
Given two $n$-point multisets $P\subseteq \mathbb{R}^2$ and
$Q\subseteq \mathbb{R}^2$, a positive integer $k$, two linear
constraints $L_1$ and $L_2$, and a linear objective function
$f:\mathbb{R}^{2}\rightarrow\mathbb{R}$,  the $k^{th}$ largest
objective value among all objective values of points in $(P\oplus
Q)_{L_1,L_2}$ can be found in $O(n\log^2 n)$ time.
\end{theorem}

\begin{theorem}\label{randomized_time}
For linear objective functions, the \textsc{Minkowski Sum Selection}
problem with two linear constraints can be solved in expected
$O(n\log n)$ time.
\end{theorem}
\begin{proof}
Due to the space limitation, we leave the proof to Appendix D.
\end{proof}

\section{Minkowski Sum Selection with $\lambda > 2$ Constraints}
Let $\lambda$ be a fixed integer greater than two. In the following
theorem, we summarize our results of the \textsc{Minkowski Sum
Selection} for the case where $\lambda$ linear constraints are given
and the objective function is linear. Due to the space limitation,
we leave the proof to Appendix~E.

\begin{theorem}\label{lambda}
Let $\lambda$ be any fixed integer larger than two. The
\textsc{Minkowski Sum Selection} problem with $\lambda$ constraints
and a linear objective function is asymptotically equivalent to the
\textsc{Minkowski Sum Selection} problem with two linear constraints
and a linear objective function.
\end{theorem}

\section{Minkowski Sum Finding}
\label{sec:Minkowski Sum Finding} In the \textsc{Minkowski Sum
Finding} problem, given two $n$-point multisets $P,Q$, a set of
$\lambda$ inequalities $Ar\geq b$, an objective function $f(x,y)$
and a real number $\delta$, we are required to find a point
$v^*=(x^*,y^*)$ among all points in $(P\oplus Q)_{Ar\geq b}$ which
minimizes $|f(x^*,y^*) - \delta|$. In this section, we show how to
cope with an objective function of the form $f(x,y)=ax+by$ or
$f(x,y)=\frac{by}{ax}$ based on the algorithms proposed by
Bernholt~\textit{et al.}~\cite{Bern}. Instead of finding the point
$v^*=(x^*,y^*)$, we would like to focus on computing the value of
$|f(x^*,y^*) - \delta|$. The point $v^*=(x^*,y^*)$ can be easily
constructed from the computed information. Before moving on to the
algorithm, let us look at the lower bound of the problem.

\begin{lemma}\label{MinkSumFindingLB}
The \textsc{Minkowski Sum Finding} problem with an objective
function of the form $f(x,y)=ax+by$ or $f(x,y)=\frac{by}{ax}$ has a
lower bound of $\Omega(n \log n)$ in the algebraic decision tree
model.
\end{lemma}
\begin{proof}
Given two real number sets $A=\{a_1, \ldots, a_n\}$ and $B=\{b_1,
\ldots, b_n\}$, the \textsc{Set Disjointness} problem is to
determine whether or not $A\cap B = \emptyset$. It is known the
\textsc{Set Disjointness} problem has a lower bound of $\Omega(n
\log n)$ in the algebraic decision tree model~\cite{Ben-Or}. We
first prove that the \textsc{Set Disjointness} problem is
linear-time reducible to the \textsc{Minkowski Sum Finding} problem
with the objective function $f(x,y) = y$. Let $P=\{(1, a_1), \ldots,
(1, a_n) \}$, $Q=\{(1, -b_1), \ldots, (1, -b_n) \}$, and $(x^*,y^*)$
be the point in $(P\oplus Q)$ such that $y^*$ is closest to zero.
Then $(x^*,y^*)=(2,0)$ if and only if $A\cap B \neq \emptyset$.
Similarly, we can prove that the \textsc{Set Disjointness} problem
is linear-time reducible to the \textsc{Minkowski Sum Finding}
problem with the objective function $f(x,y)=\frac{y}{x}$. Let
$P=\{(1, a_1), \ldots, (1, a_n) \}$, $Q=\{(1, -b_1), \ldots, (1,
-b_n) \}$, and $(x^*,y^*)$ be the point in $(P\oplus Q)$ such that
$\frac{y^*}{x^*}$ is closest to zero. Then $(x^*,y^*)=(2,0)$ if and
only if $A\cap B \neq \emptyset$.
\end{proof}

Now, let us look at how to cope with a linear objective function $f(x,y) = ax+by$.
Without loss of generality, we assume $\delta = 0$; otherwise we may perform some input transformations first.  Thus, the goal is to compute the value of $\min\{|ax+by|: (x,y)\in (P\oplus Q)_{Ar\geq b}\}$.

\begin{lemma}\label{LinearFinding}
Divide the $xy$-plane into two parts: $D_1=\{(x,y):ax+by \geq 0\}$ and $D_2=\{(x,y):ax+by < 0\}$. Given two points $v_1=(x_1,y_1)$ and $v_2=(x_2,y_2)$ in the same part, let $v_\gamma=(x_\gamma, y_\gamma) =
\gamma v_1 + (1 - \gamma) v_2$, where $\gamma \in [0,1]$. Then we
have $|a x_\gamma + b y_\gamma| \geq \min{(|a x_1 + b y_1|,
|a x_2 + b y_2|)}$.
\end{lemma}
\begin{proof}
It is easy to see the lemma holds if $b=0$. Without loss of
generality, let $x_1 \geq x_2$ and $b \neq 0$. We only prove the
case where both $v_1$ and $v_2$ are in $D_1$, and the other case can
be proved in a similar way. Now consider the following two
situations: (1) $|a x_1 + b y_1| \leq |a x_2 + b y_2|$ and (2) $|a
x_1 + b y_1| > |a x_2 + b y_2|$. In the first situation, by $|a x_1
+ b y_1| \leq |a x_2 + b y_2|$, $a x_1+b y_1 \geq 0$, and $a x_2+b
y_2 \geq 0$, we can derive that $b(y_2 - y_1) \geq a(x_1 - x_2)$.
Let $v'=(x',y')$ satisfy $a x_1 + b y_1=ax'+by'$ and
$x'=x_{\gamma}=\gamma x_1+(1-\gamma)x_2$. It follows that $y'=
\frac{a}{b} (1-\gamma) (x_1 - x_2) + y_1$. By $y_{\gamma} = \gamma
y_1 + (1 - \gamma) y_2 = (1 - \gamma) (y_2 - y_1) + y_1$, we have
$by_{\gamma} \geq by'$. Thus, $|a x_\gamma + b y_\gamma| \geq |ax'+
by'| = |a x_1 + b y_1|$. In the second situation, $b(y_1 - y_2) >
a(x_2 - x_1)$. Let $v''=(x'',y'')$ satisfy $ax_2 + by_2=ax''+by''$
and $x''=x_{\gamma}=\gamma x_1+(1-\gamma)x_2$. It follows that
$y''=\frac{a}{b} \gamma (x_2 - x_1)+y_2$. By $y_{\gamma} = \gamma
y_1 + (1 - \gamma) y_2 = \gamma (y_1 - y_2) + y_2$, we have
$by_{\gamma} > by''$. Thus, $|a x_\gamma + b y_\gamma| >
|ax''+by''|=|a x_2 + b y_2|$. Therefore, $|a x_\gamma + b y_\gamma|
\geq \min{(|a x_1 + b y_1|, |a x_2 + b y_2|)}$ if $a x_1+b y_1 \geq
0$ and $a x_2+b y_2 \geq 0$.
\end{proof}

Let $D_1=\{(x,y):ax+by \geq 0\}$ and $D_2=\{(x,y):ax+by < 0\}$. Let
$R_1$ be the vertices of the convex hull of $(P\oplus Q)_{Ar\geq b,
ax+by \geq 0}$ and $R_2$ be the vertices of the convex hull of
$(P\oplus Q)_{Ar\geq b, ax+by < 0}$. By Theorem \ref{Convexhull}, we
can compute $R_1$ and $R_2$ in $O(\lambda \log \lambda + \lambda
\cdot n \log n)$ time. Let $sol_1=\min\{|ax+by|: (x,y)\in(P\oplus
Q)_{Ar\geq b} \cap D_1\}$ and $sol_2=\min\{|ax+by|: (x,y)\in(P\oplus
Q)_{Ar\geq b} \cap D_2\}$. By Lemma \ref{LinearFinding}, we have
$sol_1=\min\{|ax+by|: (x,y) \in R_1\}$ and $sol_2=\min\{|ax+by|:
(x,y) \in R_2\}$. Note that both the sizes of  $R_1$ and $R_2$ are
bounded above by $O(\lambda\cdot n)$. Therefore, $sol_1$ and $sol_2$
are computable in $O(\lambda\cdot n)$ time by examining all points
in $R_1$ and $R_2$. Finally, we have the solution is the minimum of
$sol_1$ and $sol_2$. The total time complexity is $O(\lambda \log
\lambda + \lambda \cdot n \log n)$, and we have the following
theorem.

\begin{theorem}\label{MinkSumFindingThm}
Let $\lambda$ be any fixed nonnegative integer. The
\textsc{Minkowski Sum Finding} problem with $\lambda$ constraints
and a linear objective function can be solved in optimal $O(n\log
n)$ time.
\end{theorem}

Next, we see how to cope with an objective function of the form $f(x,y)=\frac{by}{ax}$.
Without loss of generality, we assume $\delta = 0$ and $a=b=1$; otherwise we may perform some input transformations first.  Thus, the goal is to compute the value of $\min\{|\frac{y}{x}|: (x,y)\in (P\oplus Q)_{Ar\geq b}\}$. For technical reasons, we define $\frac{y}{x}=\infty$ if $x=0$.
A function $f:D\rightarrow(\mathbb{R}\cup{\infty})$ defined on a
convex subset $D$ of $\mathbb{R}^2$ is $quasiconcave$ if whenever
$v_1,v_2\in D$ and $\gamma\in [0,1]$ then
$f(\gamma\cdot v_1+(1-\gamma)\cdot v_2) \geq \min\{f(v_2),f(v_2)\}.$

\begin{lemma}\label{SlopeFinding}
Let $D_1=\{(x,y)\in\mathbb{R}^2:x\geq 0, y \geq 0\}$,
$D_2=\{(x,y)\in\mathbb{R}^2:x\leq0, y \geq 0\}$,
$D_3=\{(x,y)\in\mathbb{R}^2:x\leq0, y \leq 0\}$, and
$D_4=\{(x,y)\in\mathbb{R}^2:x\geq0, y \leq 0\}$. Define function
$f_i: D_i \rightarrow\mathbb{R}$ by letting $f_i(x,y) =
|\frac{y}{x}|$ for each $i=1,2,3,4.$ Then we have function $f_i$ is
quasiconcave for each $i=1,2,3,4$.
\end{lemma}
\begin{proof}
We only prove that $f_1$ is quasiconcave. The proofs for
$f_2$, $f_3$, and $f_4$ can be derived in a similar way. Let $v_1=(x_1,y_1)\in D_1$,
$v_2=(x_2,y_2)\in D_1$, and $x_1\geq x_2$. Without loss of
generality we may assume $x_1>0$ and $v_\gamma \not\in\{v_1,v_2\}.$
Consider the following two cases.

Case~1: $\frac{y_1}{x_1} \leq {\frac{y_2}{x_2}}$. Let $v'=(x',y')$
be the point which satisfies $\frac{y_1}{x_1}=\frac{y'}{x'}$ and
$x'=x_{\gamma}=\gamma x_1+(1-\gamma)x_2$. By $x_1 > 0$, $x_2\geq0$,
and ${\frac{y_1}{x_1}} \leq {\frac{y_2}{x_2}}$, we have $y_2 \geq
\frac{x_2}{x_1}y_1$. It follows that $y'=(\gamma
x_1+(1-\gamma)x_2)\frac{y_1}{x_1}=\gamma y_1+(1-\gamma)
\frac{x_2}{x_1}y_1\leq \gamma y_1+(1 - \gamma)y_2=y_{\gamma}$. By $0
< x' = x_\gamma$ and $0\leq y'\leq y_\gamma$, we have
$|\frac{y_{\gamma}}{x_{\gamma}}|= \frac{y_{\gamma}}{x_{\gamma}} \geq
\frac{y'}{x'}=\frac{y_1}{x_1}=|\frac{y_1}{x_1}|\geq
\min\{|{\frac{y_1}{x_1}}|, |{\frac{y_2}{x_2}}|\}$.

Case~2: ${\frac{y_1}{x_1}} > {\frac{y_2}{x_2}}$. Let $v''=(x'',y'')$
be the point which satisfies $\frac{y_2}{x_2}=\frac{y''}{x''}$ and
$x''=x_{\gamma}=\gamma x_1+(1-\gamma)x_2$. By $x_1 > 0$ and
${\frac{y_1}{x_1}} > {\frac{y_2}{x_2}}$, we have $y_1
> \frac{x_1}{x_2}y_2$. It follows that $y''=(\gamma
x_1+(1-\gamma)x_2)\frac{y_2}{x_2}=\gamma \frac{x_1}{x_2}
 y_2+(1-\gamma) y_2 < \gamma y_1 +
(1-\gamma)y_2=y_{\gamma}$. By $0 < x'' = x_\gamma$ and $0\leq
y''\leq y_\gamma$, we have
$|\frac{y_{\gamma}}{x_{\gamma}}|=\frac{y_{\gamma}}{x_{\gamma}} >
\frac{y''}{x''}=\frac{y_2}{x_2}=|{\frac{y_2}{x_2}}|\geq
\min\{|{\frac{y_1}{x_1}}|, |{\frac{y_2}{x_2}}|\}$.
\end{proof}

Let $D_1=\{(x,y)\in\mathbb{R}^2:x\geq 0, y \geq
0\}$, $D_2=\{(x,y)\in\mathbb{R}^2:x\leq0, y \geq 0\}$,
$D_3=\{(x,y)\in\mathbb{R}^2:x\leq0, y \leq 0\}$, and
$D_4=\{(x,y)\in\mathbb{R}^2:x\geq0, y \leq 0\}$.
Let $R_i$ be the vertices of the convex hull of $(P\oplus Q)_{Ar\geq b} \cap D_i$ for $i =1,2,3,4$. By Theorem~\ref{Convexhull}, each $R_i$ is computable
in
$O(\lambda \log \lambda + \lambda \cdot n \log n)$ time. Let
$sol_i=\min\{|\frac{y}{x}|: (x,y)\in(P\oplus Q)_{Ar\geq b} \cap
D_i\}$ for each $i=1,2,3,4$. By Lemma \ref{SlopeFinding}, we have
$sol_i=\min\{|\frac{y}{x}|: (x,y) \in R_i\}$ for each $i$. Note that
the size of each $R_i$ is bounded above by $O(\lambda + n)$.
Therefore, each $sol_i$ is computable in $O(\lambda + n)$ time by
examining all points in $R_i$. Finally, we have the solution is
$\min_{i=1}^4sol_i$. The total time complexity is $O(\lambda \log \lambda + \lambda \cdot n \log n)$, and we have the following theorem.

\begin{theorem}\label{MinkDensityFindingThm}
Let $\lambda$ be any fixed nonnegative integer. The
\textsc{Minkowski Sum Finding} problem with $\lambda$ constraints
and an objective function of the form $f(x,y)=\frac{by}{ax}$ can be
solved in optimal $O(n\log n)$ time.
\end{theorem}

\section*{Acknowledgments}
We thank the anonymous referees for helpful suggestions. Cheng-Wei
Luo, Hsiao-Fei Liu, Peng-An Chen, and Kun-Mao Chao were supported in
part by NSC grants 95-2221-E-002-126-MY3 and NSC
97-2221-E-002-007-MY3 from the National Science Council, Taiwan.




\newpage

\section*{Appendix A: Applications to the Length-constrained Sum Selection Problem}
Given a sequence $S=(s_1,s_2,\ldots,s_n)$ of $n$ real numbers, and
two positive integers $l$, $u$ with $l< u$, define the length and
sum of a segment $S[i,j]=(s_i,\ldots,s_j)$ to be $length(i,j)=j-i+1$
and $sum(i,j)=\sum_{h=i}^js_h$, respectively. A segment is said to
be feasible if and only if its length is in $[l,u]$. The
\textsc{Length-Constrained Sum Selection} problem is to find the
$k^{th}$ largest sum among all sums of feasible segments of $S$.

When there are no length constraints, i.e., $l=1$ and $u=n$, the
\textsc{Length-Constrained Sum Selection} problem becomes the
\textsc{Sum Selection} problem. Bengtsson and Chen~\cite{Bengtsson}
first studied the \textsc{Sum Selection} problem and gave an $O(n
\log^2 n)$-time algorithm for it. Recently, Lin and Lee provided an
$O(n \log n)$-time algorithm~\cite{TC06} for the \textsc{Sum
Selection} problem and an expected $O(n \log (u-l+1))$-time
randomized algorithm~\cite{TC05} for the \textsc{Length-Constrained
Sum Selection} problem. In the following, we show how to solve the
\textsc{Length-Constrained Sum Selection} problem in worst-case $O(n
\log (u-l+1))$ time.

\subsection*{Algorithm}
We first reduce the \textsc{Length-Constrained Sum Selection}
problem to the \textsc{Minkowski Sum Selection} problem as follows.
Let $P = \{p_0,p_1,\ldots,p_n\}$ and
$Q=\{q_0,q_{-1},\ldots,q_{-n}\}$, where
$p_i=(x_i,y_i)=(i,\sum_{t=1}^{i}s_t)$ and $q_{-i} =
(\overline{x}_i,\overline{y}_i)=(-i,-\sum_{t=1}^{i}s_t)$ for all
$i=0,1,\ldots,n$.

A point $(x,y)$ in $P\oplus Q$ is said to be a
feasible point if and only if $l\leq x \leq u$. Each feasible
segment $S[i,j]$ corresponds to a feasible point $(x,y)=p_j+q_{1-i}$
in $P\oplus Q$. Thus, the \textsc{Length-Constrained Sum Selection}
problem is equivalent to finding the $k^{th}$ largest $y$-coordinate
among all $y$-coordinates of feasible points in $P\oplus Q$. We next
show how to do this in $O(n\log (u-l+1))$ time. For simplicity, we
assume $n$ is a multiple of $u-l$.

\begin{enumerate}
\item Let $i_t = t(u-l)$ and $j_t = l-t(u-l)$ for $t = 0,1,\ldots,\frac{n}{u-l}$.
    \item For $t \leftarrow 0$ to $\frac{n}{u-l}$ do

        \begin{enumerate}
        \item Let $P_t$=$\{p_h\in P: i_t-(u-l)<h\leq i_t\}$ and $Q_t = Q_{t,1}\cup Q_{t,2}$, where $Q_{t,1}=\{q_h\in Q: j_t\leq h< j_t+(u-l)\}$ and $Q_{t,2}=\{q_h\in Q: j_t+(u-l)\leq h< j_t+2(u-l)\}.$
        \item Store the $y$-coordinates of points in $(P_t\oplus Q_{t,1})_{x\geq l}$ as a set $N_{t,1}$ of sorted matrices such that the sum of side lengths of the sorted matrices in $N_{t,1}$ is no greater than $c\cdot((|P_t|+|Q_{t,1}|)\log (|P_t|+|Q_{t,1}|+1))$ for some constant $c$.
        \item Store the $y$-coordinates of points in $(P_t\oplus Q_{t,2})_{x\leq u}$ as a set $N_{t,2}$ of sorted matrices such that the sum of side lengths of the sorted matrices in $N_{t,2}$ is no greater than $c\cdot((|P_t|+|Q_{t,2}|)\log (|P_t|+|Q_{t,2}|+1))$ for some constant $c$.\\
        \end{enumerate}
    \vspace{-12pt}
    \item Return the $k^{th}$ largest element among the elements of sorted matrices in
    $\bigcup_{t=0}^{\frac{n}{(u-l)}}(N_{t,1}\cup N_{t,2})$.
  \end{enumerate}

The following lemma ensures the correctness.

\begin{lemma}\label{parallel_correct}
$(P\oplus Q)_{l\leq x \leq u}
=\bigcup_{t=0}^{\frac{n}{(u-l)}}((P_t\oplus Q_{t,1})_{l\leq
x}\cup(P_t\oplus Q_{t,1})_{x\leq u})$.
\end{lemma}
\begin{proof}
We prove that $(P\oplus Q)_{l\leq x \leq u}\stackrel{(1)} =
\bigcup_{t=0}^{\frac{n}{(u-l)}}(P_t\oplus Q)_{l\leq x \leq
u}\stackrel{(2)}=\bigcup_{t=0}^{\frac{n}{(u-l)}}(P_t\oplus
Q_t)_{l\leq x \leq
u}\stackrel{(3)}=\bigcup_{t=0}^{\frac{n}{(u-l)}}((P_t\oplus
Q_{t,1})_{l\leq x \leq u}\cup (P_t\oplus Q_{t,2})_{l\leq x \leq
u})\stackrel{(4)}=\bigcup_{t=0}^{\frac{n}{(u-l)}}((P_t\oplus
Q_{t,1})_{l\leq x}\cup(P_t\oplus Q_{t,2})_{x\leq u}).$ It is clear
that equations (1) and (3) are true, so only equations (2) and (4)
remain to be proved.

We first prove equation (2) by showing that $(P_t\oplus Q)_{l\leq x
\leq u}=(P_t\oplus Q_t)_{l\leq x \leq u}.$ Suppose for contradiction
that there exist $p_i \in P_t$  and $q_j \not\in Q_t$ such that
$l\leq x_i+\overline{x}_j = i+j\leq u.$ By $p_i\in P_t$, we have
$(t-1)(u-l)<i\leq t(u-l)$; by $q_j \not\in Q_t$, we have either $j <
l-t(u-l)$ or $j\geq l-(t-2)(u-l)$. It follows that $i+j$ is either
less than $l$ or larger than $u$, a contradiction. To prove equation
(4), it suffices to prove that all points in $(P_t\oplus Q_{t,1})$
must have $x$-coordinates less than $u$ and all points in
$(P_t\oplus Q_{t,2})$ must have $x$-coordinates larger than $l$. Let
$p_{i}\in P_t$, $q_{j} \in Q_{t,1}$ and $q_{j'}\in Q_{t,2}$. It
follows that $(t-1)(u-l)<x_i=i\leq t(u-l)$, $l-t(u-l)\leq
\overline{x}_j = j < l-(t-1)(u-l)$, and $l-(t-1)(u-l)\leq
\overline{x}_{j'}=j' < l-(t-2)(u-l)$. Thus, we have
$x_i+\overline{x}_j = i+j<u$ and $l<x_i+\overline{x}_{j'}=i+j'$.
\end{proof}

Since $|P_t|, |Q_{t,1}|$, and $|Q_{t,2}|$ are no greater than
$(u-l)$ for all $t$, each execution of Step~2.b and Step~2.c can be
done in $O((u-l)\log(u-l+1))$ time by Lemma~\ref{IndexSum}. There
are total $\frac{n}{u-l}+1$ iterations of the for-loop in
Step~2, so the total time spent on Step~2.b and Step~2.c is $O(n
\log(u-l+1)).$ The sum of side lengths of sorted matrices in
$\bigcup_{t=0}^{\frac{n}{(u-l)}}(N_{t,1}\cup N_{t,2})$ is
$O(\sum_{t=0}^ {\frac{n}{u-l}}\sum_{i=1}^2(|P_t|+|Q_{t,i}|)\log
(|P_t|+|Q_{t,i}|+1)) = O(\sum_{t=0}^
{\frac{n}{u-l}}(u-l)\log(u-l+1)) = O(n \log (u-l+1))$. Therefore, by
Theorem~\ref{SelectionTime}, Step~3 can be done in $O(n \log
(u-l+1))$ time. Putting everything together, we have that the total
running time is $O(n \log (u-l+1))$.

\begin{theorem}
The \textsc{Length-Constrained Sum Selection} problem can be solved
in $O(n\log(u-l+1))$ time.
\end{theorem}

\section*{Appendix B: Applications to the Density Finding Problem}
Given a sequence of number pairs $S=((s_1, w_1), (s_2, w_2), \ldots,
(s_n, w_n))$ where $w_i > 0$ for $i = 1, 2, \ldots, n$, two positive
numbers $l,u$ with $l < u$, and a real number $\delta$, let
\textit{segment} $S(i,j)$ of $S$ be the consecutive subsequence of
$S$ between indices $i$ and $j$. Define the \textit{sum} $s(i,j)$,
\textit{width} $w(i,j)$, and \textit{density} $d(i,j)$ of segment
$S(i,j)$ to be $\sum_{r=i}^{j} s_r$, $\sum_{r=i}^{j} w_r$ and
$\frac{s(i,j)}{w(i,j)},$ respectively. A segment $S(i,j)$ is said to
be \textit{feasible} if and only if $l \leq w(i,j) \leq u$. The
\textsc{Density Finding} problem is to compute the density of the
feasible segment $S(i^*,j^*)$ which minimizes $|d(i^*,j^*)-\delta|$.
Lee~$et~al.$~\cite{LLL} proved that the \textsc{Density Finding}
problem has a lower bound of $\Omega(n \log n)$ in the algebraic
decision tree model and provided an $O(n\log^2 m)$ algorithm for it,
where $m = \min(\lfloor \frac{u-l}{w_{min}}\rfloor ,n)$ and $w_{min}
= \min_{1\leq r\leq n} w_r$. In the following we describe how to
solve the \textsc{Density Finding} problem in $O(n\log n)$ time by
using the algorithm developed in Section~\ref{sec:Minkowski Sum Finding}.

Let $w(1,0)=0$ and $s(1,0)=0$. Compute in $O(n)$ time the
following two point sets: $P=\{(w(1,i),s(1,i))| 0\leq i\leq n\}$ and
$Q=\{(-w(1,i),-s(1,i))|0\leq i \leq n)\}$. Note that each feasible segment
$S(i,j)$ of $S$ corresponds to a point
$(w(1,j)-w(1,i-1),s(1,j)-s(1,i-1))$ in $(P\oplus Q)_{l\leq x \leq
u}$. Thus, the problem is reduced to
finding the point $(x^*,y^*)$ in $(P\oplus Q)_{l\leq x \leq u}$ such
that $|\frac{y^*}{x^*} - \delta|$ is minimized.
By Theorem~\ref{MinkDensityFindingThm}, it can be done in $O(n\log n)$ time,
so we have the following theorem.

\begin{theorem}
The \textsc{Density Finding} problem can be solved in optimal
$O(n\log n)$ time.
\end{theorem}

\section*{Appendix C: Minkowski Sum Selection with Two Parallel Constraints}
Now we explain how to solve the \textsc{Minkowski Sum Selection}
problem with two parallel constraints in $O(n\log n)$ time. Given
$P=\{(x_{1,1},y_{1,1}),\ldots,(x_{1,n},y_{1,n})\}$,
$Q=\{(x_{2,1},y_{2,1}),\ldots,(x_{2,n},y_{2,n})\}$, a positive
integer $k$, two parallel constraints $L_1$: $ax+by\geq c_1$ and
$L_2$: $ax+by\leq c_2$ with $c_1\leq c_2$, and a linear objective
function $f(x,y)=dx+ey$, where $a$, $b$, $c_1$, $c_2$, $d$, and $e$
are all constants, we want to find the $k^{th}$ largest objective
value among all objective values of points in $(P\oplus
Q)_{L_1,L_2}$. Note that if the constraints $L_1$ and $L_2$ are of
the forms $ax+by\leq c_1$ and $ax+by\leq c_2$ respectively, this
problem degenerates to the \textsc{Minkowski Sum Selection} problem
with one constraint and can be solved by the algorithm stated in
Section~3. We perform the following transformation.

\begin{enumerate}
    \item Change the content of $P$ and $Q$ to
    $\{(ax_{1,1}+by_{1,1},dx_{1,1}+ey_{1,1}),\ldots,(ax_{1,n}+by_{1,n},dx_{1,n}+ey_{1,n})\}$,
    and
    $\{(ax_{2,1}+by_{2,1},dx_{2,1}+ey_{2,1}),\ldots,(ax_{2,n}+by_{2,n},dx_{2,n}+ey_{2,n})\}$, respectively.
    \item Change the constraints from $ax+by \geq c_1$ and $ax+by\leq c_2$ to $x\geq c_1$ and
            $x\leq c_2$, respectively.
    \item Change the objective function from $dx+ey$ to $y$.
\end{enumerate}

This transformation can be done in $O(n)$ time and the answer
remains the same. Hence from now on, our goal becomes to find the
$k^{th}$ largest $y$-coordinate on the constrained Minkowski sum of
$P$ and $Q$ subject to the constraints $L_1$: $x\geq c_1$ and $L_2$:
$x\leq c_2$. First we sort $P$ and $Q$ into $P_x$ and $Q_x$ in
nondecreasing order of $x$-coordinates, respectively in $O(n\log n)$
time. Let $P_x=\{(x_1,y_1),\ldots,(x_n,y_n)\}$ and
$Q_x=\{(\bar{x}_1,\bar{y}_1),\ldots,(\bar{x}_n,\bar{y}_n)\}$. For
all points in $P_x$, we can form a partition of them according to
the values of their $x$-coordinates. Let the partition be
$P_{t_1},P_{t_2},\ldots,P_{t_m}$ where $P_{t_i}=\{(x_j,y_j)\in
P_x:(t_i-1)(c_2-c_1)<x_j\leq t_i(c_2-c_1)\}$, and $t_i$ be an
integer for $i=1,2,\ldots,m$ with $t_1<t_2<\ldots<t_m$. Similarly,
we can partition the points in $Q_x$ according to the values of
their $x$-coordinates. Let the partition be
$Q_{\bar{t}_1},Q_{\bar{t}_2},\ldots,Q_{\bar{t}_{m'}}$ where
$Q_{\bar{t}_i}=\{(\bar{x}_j,\bar{y}_j)\in
Q_x:c_1-\bar{t}_i(c_2-c_1)\leq
\bar{x}_j<c_1-(\bar{t}_i-1)(c_2-c_1)\}$, and $\bar{t}_i$ be an
integer for $i=1,2,\ldots,m'$ with
$\bar{t}_1<\bar{t}_2<\ldots<\bar{t}_{m'}$. Since $P_x$ and $Q_x$ are
sorted in nondecreasing order of $x$-coordinates respectively, the
two partitions can be easily produced in linear time. In the
following, we show the algorithm for this problem.

\begin{enumerate}
\item Let $P_{t_1},P_{t_2},\ldots,P_{t_m}$ and $Q_{\bar{t}_1},Q_{\bar{t}_2},\ldots,Q_{\bar{t}_{m'}}$
        be defined as the above.
    \item For $i \leftarrow 1$ to $m$ do
        \begin{enumerate}\itemsep=-1pt
        \item To find $Q_{t_i}=\{(\bar{x}_j,\bar{y}_j)\in Q_x: c_1-t_i(c_2-c_1)\leq \bar{x}_j< c_1-(t_i-1)(c_2-c_1)\}$
                and $Q_{t_i-1}=\{(\bar{x}_j,\bar{y}_j)\in Q_x: c_1-(t_i-1)(c_2-c_1)\leq \bar{x}_j< c_1-(t_i-2)(c_2-c_1)\}.$
        \item If $Q_{t_i}$ exists, store the $y$-coordinates of points in $(P_{t_i}\oplus Q_{t_i})_{x\geq c_1}$
                as a set $N_{i,1}$ of sorted matrices such that the sum of side lengths of the sorted
                matrices in $N_{i,1}$ is no greater than $c\cdot((|P_{t_i}|+|Q_{t_i}|)\log (|P_{t_i}|+|Q_{t_i}|+1))$
                for some constant $c$.
        \item If $Q_{t_i-1}$ exists, store the $y$-coordinates of points in $(P_{t_i}\oplus Q_{t_i-1})_{x\leq c_2}$
                as a set $N_{i,2}$ of sorted matrices such that the sum of side lengths of the sorted
                matrices in $N_{i,2}$ is no greater than $c\cdot((|P_{t_i}|+|Q_{t_i-1}|)\log (|P_{t_i}|+|Q_{t_i-1}|+1))$
                for some constant $c$.\\
        \end{enumerate}
    \vspace{-15pt}
    \item Return the $k^{th}$ largest element among the elements of sorted matrices in
    $\bigcup_{i=1}^{m}(N_{i,1}\cup N_{i,2})$.
  \end{enumerate}

By the proof of Lemma~\ref{parallel_correct}, we ensure the
correctness of the algorithm. Now we consider the time complexity of
the algorithm. By Lemma~\ref{IndexSum}, each execution of Step~2.b
and Step~2.c can be done in $O((|P_{t_i}|+|Q_{t_i}|)\log
(|P_{t_i}|+|Q_{t_i}|+1))$ and $O((|P_{t_i}|+|Q_{t_i-1}|)\log
(|P_{t_i}|+|Q_{t_i-1}|+1))$ time, respectively. Since there are
total $m$ iterations of the for-loop in Step~2, it follows
that

\begin{eqnarray*}
& &\sum_{i=1}^{m}\big((|P_{t_i}|+|Q_{t_i}|)\log
(|P_{t_i}|+|Q_{t_i}|+1)+(|P_{t_i}|+|Q_{t_i-1}|)\log
(|P_{t_i}|+|Q_{t_i-1}|+1)\big)\\&\leq&
\sum_{i=1}^{m}\big((|P_{t_i}|+|Q_{t_i}|)\log
(2n+1)+(|P_{t_i}|+|Q_{t_i-1}|)\log(2n+1)\big)\\&\leq&
\sum_{i=1}^{m}(2|P_{t_i}|+|Q_{t_i}|+|Q_{t_i-1}|)\log(2n+1)\leq
(4n)\log(2n+1).
\end{eqnarray*}

Therefore, the total time spent on Step~2 is $O(n\log n)$ and the
sum of the side lengths of sorted matrices in
$\bigcup_{i=1}^{m}(N_{i,1}\cup N_{i,2})$ is also $O(n\log n)$. By
Theorem~\ref{SelectionTime}, Step~3 can be done in $O(n\log n)$
time. Putting everything together, we have that the total running
time is $O(n\log n)$. The next theorem summarizes the time
complexity of the algorithm.

\begin{theorem}
For linear objective functions, the \textsc{Minkowski Sum Selection}
problem with two parallel constraints can be solved in $O(n\log n)$
time.
\end{theorem}

\section*{Appendix D: A Randomized Algorithm for Minkowski Sum Selection with
Two Constraints}In this section, we introduce a randomized algorithm
for the \textsc{Minkowski Sum Selection} problem with two
constraints that runs in expected $O(n\log n)$ time.

\subsection*{Subroutines for Minkowski Sum Selection Problem with Two
Constraints} Our randomized algorithm is based on three subroutines
for three subproblems. In this subsection, we define these
subproblems and give these subroutines for them.

Before we discuss these subroutines, we introduce the notion of an
\textit{order-statistic tree}. An order-statistic tree is a balanced
search tree with additional information, $size[z]$, stored in each
node $z$ of the tree. The additional information $size[z]$ contains
the total number of nodes in the subtree rooted at $z$. Define
$left[z]$ and $right[z]$ are the left and right children of the node
$z$, respectively. The additional information $size[z]$ equals to
$size[left[z]]+size[right[z]]+1$ if $z$ is an internal node, and one
if $z$ is a leaf node. Let $key[z]$ be the key of the node $z$. The
rank of a given value $x$ can be determined in $O(\log n)$ time by
using the order-statistic tree $T$, where $n$ is the number of nodes
in $T$. That is, we can find the rank $r(x,T)=|\{y|y\in
T,key[y]>x\}|$ in $O(\log n)$ time, retrieve an element in $T$ with
a given rank in $O(\log n)$ time and maintain both insertion and
deletion operations in $T$ in $O(\log n)$ time.

The first subproblem is the reporting version of the
\textsc{Minkowski Sum Range Query} problem with two constraints,
which is defined as follows: Given two $n$-point multisets $P$, $Q$,
two constraints $L_1$, $L_2$, and two real numbers $s_l$, $s_r$ with
$s_l\leq s_r$, we want to output all points in $(P\oplus
Q)_{L_1,L_2}$, such that their $y$-coordinates are in the range
$[s_l,s_r]$. Before we discuss this subproblem, we consider a weak
version of this subproblem. The weak version is defined as above,
except that we assign the constraints $L_1$: $ax+by\geq c_1$ and
$L_2$: $ax+by\leq c_2$, where $c_1\leq c_2$. To solve the weak
version, we perform the input transformation stated in Section~4.1.1
to change $L_1$ and $L_2$ to $x\geq c_1$ and $x\leq c_2$
respectively. Then we sort $P$ and $Q$ into $P_x$ and $Q_x$
respectively in nondecreasing order of $x$-coordinates. Let
$P_x=((x_1,y_1),\ldots,(x_n,y_n))$ and
$Q_x=((\bar{x}_1,\bar{y}_1),\ldots,(\bar{x}_n,\bar{y}_n))$. Denote
by $l_i$ the smallest index in $Q_x$ such that
$x_i+\bar{x}_{l_i}\geq c_1$ and by $r_i$ the largest index in $Q_x$
such that $x_i+\bar{x}_{r_i}\leq c_2$. It is guaranteed that
$l_i\geq l_j$ and $r_i\geq r_j$ for any $i<j$. For this reason, it
can be easily done to find all $l_i$ and $r_i$ for $i=1,\ldots,n$ in
total $O(n)$ time. We use the points in $Q_x$ to construct an
order-statistic tree with the values of $y$-coordinates as keys. For
the index $j$, we use the points $(\bar{x}_{i},\bar{y}_{i})$ in
$Q_x$ for $i=l_j,\ldots,r_j$ to construct an order-statistic tree
$T(j)$ in $O((r_j-l_j)\log(r_j-l_j))$ time. Because $T(j)$ is also a
balanced binary search tree, we can report all points in $T(j)$
whose $y$-coordinates are in $[s_l-y_j,s_r-y_j]$ by binary search in
$O(log(r_j-l_j)+h_j)$ time, where $h_j$ is the total number of
points whose $y$-coordinates are in $[s_l-y_j,s_r-y_j]$. At each
iteration $j$, we can maintain $T(j+1)$ dynamically by deleting all
points $(\bar{x}_i,\bar{y}_i)$ in $T(j)$ for
$i=r_{j+1}+1,\ldots,r_j$ and inserting all points
$(\bar{x}_{i'},\bar{y}_{i'})$ into $T(j)$ for
$i'=l_{j+1},\ldots,l_j-1$. It suffices to iterate on each index $j$
to find all points in $(P\oplus Q)_{L_1,L_2,y\leq s_r,y\geq s_l}$.
Hence, we can solve the weak version in $O(n\log n+h)$ time, where
$h$ is the output size.

\begin{figure}
\centering
\includegraphics[scale=0.6]{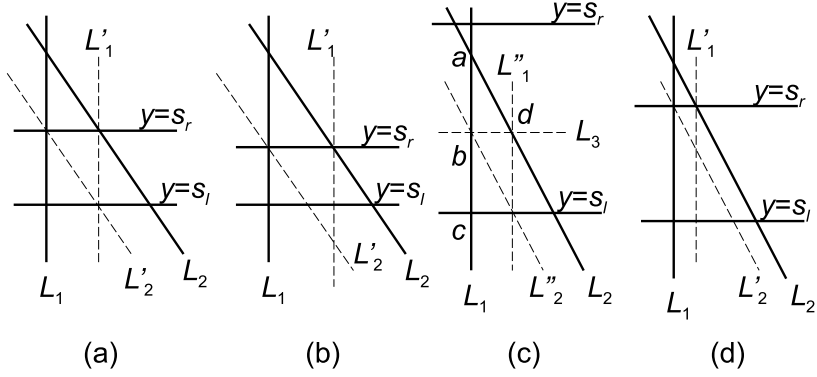}
\caption{The illustration of how to solve the reporting version.}
\label{reporting}
\end{figure}

Now we show how to solve the reporting version of the
\textsc{Minkowski Sum Range Query} problem with two constraints by
the weak version stated above. By performing the transformation
stated in Section~4.1.1, $L_1$ and $L_2$ are changed to $x\geq c_1$
and $ax+by\geq c_2$ respectively. We divide the feasible region
bounded by $L_1$, $L_2$, and $s_l\leq y\leq s_r$ into several
subregions. For each subregion, it can be solved by the subroutine
for the weak version of this subproblem. Let $L_1'$ be the line that
is parallel to $L_1$ and passes through the intersection point of
$L_2$ and $y=s_r$, and $L_2'$ be the line that is parallel to $L_2$
and passes through the intersection point of $L_1$ and $y=s_r$. By
the location of the intersection point of $L_1'$ and $L_2'$, we have
four possible cases: (a) the intersection point lies in the line
$y=s_l$; (b) the intersection point lies below the line $y=s_l$; (c)
the intersection point lies above the line $y=s_r$; (d) the
intersection point lies between $y=s_l$ and $y=s_r$. See
Figure~\ref{reporting} for an illustration. We solve the reporting
version according to the four possible cases respectively. For
Case~(a), we consider the parallelogram formed by $L_2$, $L_2'$,
$y\leq s_r$, and $y\geq s_l$ and all feasible points in this
parallelogram can be reported by the subroutine of the weak version.
Next we consider the rectangle formed by $L_1$, $L_1'$, $y\leq s_r$,
and $y\geq s_l$ and all feasible points in this rectangle can be
reported in the same way, except we have to remove the redundant
points in the area formed by $L_1'$, $L_2'$, and $y\leq s_r$. When
we report each feasible point in the rectangle formed by $L_1$,
$L_1'$, $y\leq s_r$, and $y\geq s_l$, we can check whether this
point lies in the area formed by $L_1'$, $L_2'$, and $y\leq s_r$ in
$O(1)$ time. If this point lies in this area, we discard this point,
or we report it. For Case~(b), we can also solve this case in the
same way of Case~(a), except the redundant points are in the area
formed by $L_1'$, $L_2'$, $y\leq s_r$, and $y\geq s_l$. For each
redundant point, however, the removal can be also easily done in
$O(1)$ time.

For Case~(c), we must divide the triangle formed by $L_1$, $L_2$,
and $y\geq s_l$ in another way. Let $a$ be the intersection point of
$L_1$ and $L_2$, $c$ be the intersection point of $L_1$ and $y\geq
s_l$, and $b$ be the middle point of the line segment
$\overline{ac}$. We can draw the line $L_3$ that is parallel to
$y=s_l$ and passes through $b$, and let $d$ be the intersection
point of $L_3$ and $L_2$. Let $L_2''$ be the line that is parallel
to $L_2$ and passes through $b$, and $L_1''$ be the line that is
parallel to $L_1$ and passes through $d$. Because the triangle
formed by $L_1$, $L_2$, and $y\geq s_l$ is a right-angled triangle,
the intersection point of $L_1''$ and $L_2''$ must lie in the line
$y=s_l$. We first report the feasible points in the parallelogram
formed by $L_1$, $L_1''$, $L_2$, and $L_2''$ by the subroutine of
the weak version. Then we report the feasible points in the
parallelogram formed by $L_3$, $y\geq s_l$, $L_2$, and $L_2''$ in
the same way and remove the redundant points, i.e., the points lie
in the area formed by $L_3$, $L_1''$, and $L_2''$. Finally, we
report the feasible points in the rectangle formed by $L_1$,
$L_1''$, $L_3$, and $y\geq s_l$ and remove the redundant points in
the area formed by $L_3$, $L_1''$, and $L_2''$.

For Case~(d), we report all feasible points in the parallelogram
formed by $L_2$, $L_2'$, $y\leq s_r$, and $y\geq s_l$ and in the
rectangle formed by $L_1$, $L_1'$, $y\leq s_r$, and $y\geq s_l$ with
the removal of the redundant points in the area formed by $L_1'$,
$L_2'$, and $y\leq s_r$. The remaining is the triangle formed by
$L_1'$, $L_2'$, and $y\geq s_l$ and is just Case~(c). Therefore, we
can solve this triangle in the same way of Case~(c). In the
following, we conclude the time complexity of this problem.

\begin{lemma}
The reporting version of the \textsc{Minkowski Sum Range Query}
problem with two constraints can be solved in $O(n\log n+h)$ time,
where $h$ is the output size.
\end{lemma}

The second subproblem, called the counting version of the
\textsc{Minkowski Sum Range Query} problem with two constraints, is
defined as before, but we only want to find the number of points in
$(P\oplus Q)_{L_1,L_2}$ satisfying the range query, i.e., their
$y$-coordinates are between $s_l$ and $s_r$. To solve this
subproblem, we make use of the procedure Ranking$_2$ shown in
Figure~\ref{algo4}. Let $R_l$ be the number of points in $(P\oplus
Q)_{L_1,L_2}$ that their $y$-coordinates are larger than $s_l$. It
is obvious to see that $R_l$ is equal to
Ranking$_2(P,Q,L_1,L_2,f(x,y)=y,s_l)$. Let $R_r$ be the number of
points in $(P\oplus Q)_{L_1,L_2}$ that their $y$-coordinates are
larger than $s_r$, and $R_r$ is equal to
Ranking$_2(P,Q,L_1,L_2,f(x,y)=y,s_r)$. As a result, the number of
points in $(P\oplus Q)_{L_1,L_2,y\leq s_r,y\geq s_l}$ is $R_l-R_r$
plus the number of points in $(P\oplus Q)_{L_1,L_2,y=s_l}$. Let $R$
be the number of points in $(P\oplus Q)_{L_1,L_2,y=s_l}$. $R$ can be
obtained by computing Ranking$_1(P,Q,y=s_l,f(x,y)=x,c_1)$ minus
Ranking$_1(P,Q,y=s_l,f(x,y)=x,c_2)$. Hence, we obtain the number of
points in $(P\oplus Q)_{L_1,L_2,y\leq s_r,y\geq s_l}$.

\begin{lemma}
The counting version of the \textsc{Minkowski Sum Range Query}
problem with two constraints can be solved in $O(n\log n)$ time.
\end{lemma}

The last subproblem, called the \textsc{Random Sampling Minkowski
Sum} problem with two constraints, is defined as follows: Give two
$n$-point multisets $P$, $Q$, two constraints $L_1$: $x\geq c_1$,
$L_2$: $ax+by\geq c_2$, and two real numbers $s_l$, $s_r$ with
$s_l\leq s_r$, we want to randomly generate $n$ points from
$(P\oplus Q)_{L_1,L_2,y\leq s_r,y\geq s_l}$ with replacement. For
the ease of similar discussions, in the following we only focus on
Case~(a) illustrated in Figure~\ref{reporting} and the other cases
can be solved in a similar way.

Let $N$ be the number of points in $(P\oplus Q)_{L_1,L_2,y\leq
s_r,y\geq s_l}$, $N_1$ be the number of points in the parallelogram
$A$ formed by $L_2$, $L_2'$, $y\leq s_r$ and $y\geq s_l$, $N_2$ be
the number of points in the rectangle $B$ formed by $L_1$, $L_1'$,
$y\leq s_r$ and $y\geq s_l$, and $N_3$ be the number of points in
the triangle $C$ formed by $L_1'$, $L_2'$ and $y\leq s_r$. $N$,
$N_1$, $N_2$, and $N_3$ can be computed by the subroutine for the
counting version of the \textsc{Minkowski Sum Range Query} problem
with two constraints. For the parallelogram $A$, we can use the
subroutine for the weak version of the reporting subproblem to
construct the order-statistic tree $T(j)$ on points in $Q_x$ for
$j=1,\ldots,n$. Let the size of $T(j)$ be $t_j$ and then
$N_1=t_1+t_2+\ldots+t_n$. We first pick $n$ random integers
$\bar{R}=\{\bar{r}_1,\bar{r}_2,\ldots,\bar{r}_n\}$ uniformly
distributed in the range from $1$ to $N$ with replacement. Since
$N=O(n^2)$, we can sort them by radix sort and rename them such that
$\bar{r}_1\leq \bar{r}_2\leq\ldots\leq \bar{r}_n$ in $O(n)$ time.
Let $\tau_j=t_1+t_2+\ldots+t_j$. For each $j$, there exist
$\bar{r}_c,\bar{r}_{c+1},\ldots,\bar{r}_{c+d}\in\bar{R}$ such that
$\tau_{j-1}<\bar{r}_c\leq \bar{r}_{c+1}\leq\ldots\leq
\bar{r}_{c+d}\leq\tau_j$. For $\bar{r}_i\leq N_1$, we shall find
points $\bar{s}_c,\bar{s}_{c+1},\ldots,\bar{s}_{c+d}$ in $(P\oplus
Q)_{L_1,L_2,y\leq s_r,y\geq s_l}$ with a one-to-one correspondence
to $\bar{r}_c,\bar{r}_{c+1},\ldots,\bar{r}_{c+d}$. For each index
$j$, we make a query to the order-statistic tree $T(j)$ in order to
count the total number $\alpha_j$ of points such that their
$y$-coordinates are less than $s_l$, i.e., $\alpha_j=|\{y|y\in
T(j),key[y]<s_l\}|$. We then retrieve the point
$\bar{s}_{c+i}=(\bar{x}_{q_i},\bar{y}_{q_i})$ from $T(j)$ such that
$\bar{y}_{q_i}$ has a rank equal to
$\alpha_j+\bar{r}_{c+i}-\tau_{j-1}$ in $T(j)$ for each
$i=0,1,\ldots,d$. For $\bar{r}_i>N_1$, we record the total number of
$\bar{r}_i$ larger than $N_1$, say $\phi$. We thus obtain a set of
points, $\bar{S}=\{\bar{s}_1,\ldots,\bar{s}_{N_1}\}$. Next we remove
the points in $\bar{S}$ lying in the triangle $C$, and record the
number of points removed, say $\psi$. Then we select $\phi+\psi$
points randomly from the rectangle $B$ using the same way. Combining
these $\phi+\psi$ points with $\bar{S}$, we obtain the set of random
sampling points.

Now we show that the sampling resulted from our random sampling
subroutine is uniformly random. Let $\bar{X}=\{X_1,X_2,\ldots,X_n\}$
be any fixed random sample generated from our random sampling
subroutine and $X_{i_1},X_{i_2},\ldots X_{i_k}$ be the points in the
triangle formed by $L_1'$, $L_2$, and $y\geq s_l$. The probability
that this random sample occurs is exactly
$(\frac{1}{N})^k(\frac{N_2}{N})^{n-k}(\frac{1}{N_2})^{n-k}=(\frac{1}{N})^n$.
Thus we obtain that for any fixed random sample, the probability of
occurring is the same $(\frac{1}{N})^n$, i.e., the random sample
generated from our random sampling subroutine is uniformly random.
The following lemma concludes the time complexity.

\begin{lemma}
The \textsc{Random Sampling Minkowski Sum} problem with two
constraints can be solved in expected $O(n\log n)$ time.
\end{lemma}

\subsection*{Algorithm} First of all, we perform the input
transformation stated in Section~4.1.1. Hence we assume that the two
constraints are $L_1$: $x\geq c_1$, $L_2$: $ax+by\geq c_2$ and the
objective function is $f(x,y)=y$. We say a range $[s_l,s_r]$
\textit{contains} a point $s \in P\oplus Q$ if $s$ satisfies two
linear constraints $y \geq s_l$ and $y \leq s_r$.

Let the $y$-coordinates of the points in $P\oplus Q$ be in the range
$[s_l,s_r]$, and $s^*=(x^*,y^*)$ be the solution of the
\textsc{Minkowski Sum Selection} problem with two constraints. Our
randomized algorithm for this problem will contract the range
$[s_l,s_r]$ into a smaller range $[s_{l'},s_{r'}]$ such that
$[s_{l'},s_{r'}]$ contains $s^*$. A point $s$ is said to be
\textit{feasible} if $s \in (P\oplus Q)_{L_1,L_2}$. Let $N$ be the
number of points in $(P\oplus Q)_{L_1,L_2}$. The subrange
$[s_{l'},s_{r'}]$ will contain at most $O(N/\sqrt{n})$ feasible
points. We shall repeat the contraction several times until the
subrange contains at most $O(n)$ feasible points and also the
solution $s^*$. Then we output all feasible points in this subrange
by the subroutine for the reporting version of the \textsc{Minkowski
Sum Range Query} problem with two constraints and find the solution
$s^*$ with an appropriate rank.

We first randomly select $n$ independent feasible points
$\bar{S}=\{\bar{s}_1,\bar{s}_2,\ldots,\bar{s}_n\}$ which are
contained in the range $[s_l,s_r]$ from $N$ feasible points by the
subroutine for the \textsc{Random Sampling Minkowski Sum} problem in
$O(n\log n)$ time. When we randomly select a feasible point in
$[s_l,s_r]$, the probability is $\frac{k}{N}$ that its
$y$-coordinate is smaller than that of $s^*$. Consider this event as
a Bernoulli trial with the success probability $p=\frac{k}{N}$. It
is obvious to see that the total number of successes is a random
variable which has a binomial distribution. Hence the expected value
of the total number of successes is $\mu=np=n\frac{k}{N}$. As a
result, we know the good approximation for the point of the $k^{th}$
largest $y$-coordinate among all feasible points is the point of the
$e^{th}$ largest $y$-coordinate among $\bar{S}$, where $e=\lfloor
np\rfloor=\lfloor n\frac{k}{N}\rfloor$.

Let $l'=\max\{1,\lfloor n\frac{k}{N}-t\frac{\sqrt{n}}{2}\rfloor\}$
and $r'=\min\{n,\lfloor n\frac{k}{N}+t\frac{\sqrt{n}}{2}\rfloor\}$,
where $t$ is a constant and will be determined later. After the
random sampling, we can find the $l'$-th and the $r'$-th largest
$y$-coordinates in $\bar{S}$, say $s_{l'}$ and $s_{r'}$
respectively, by any standard selection algorithm in $O(n)$ time to
obtain the subrange $[s_{l'},s_{r'}]$. Next, we check the following
two conditions by the subroutine for the counting version of the
\textsc{Minkowski Sum Range Query} problem with two constraints in
$O(n\log n)$ time:

(1) The subrange $[s_{l'},s_{r'}]$ contains the solution $s^*$.

(2) The subrange $[s_{l'},s_{r'}]$ contains at most
$t^2N/(t-1)\sqrt{n}$ $(<2tN/\sqrt{n})$ feasible points.

Let $k_1$ and $k_2$ be the total number of feasible points contained
in $[s_l,s_{l'})$ and $[s_l,s_{r'}]$ respectively. If $s^*$ is
contained in the subrange $[s_{l'},s_{r'}]$, we know that $k_1<k$
and $k_2\geq k$. If both of the two conditions hold, we replace the
range $[s_l,s_r]$ and the rank $k$ with the subrange
$[s_{l'},s_{r'}]$ and the rank $k'=k-k_1$. If any of the two
conditions is violated, we repeat the above step until both of the
two conditions are satisfied, i.e. we need to select $n$ random
feasible points with replacement in the range $[s_l,s_r]$ by running
the subroutine for the \textsc{Random Sampling Minkowski Sum}
problem again to obtain a new subrange $[s_{l'},s_{r'}]$ and then
check the above two conditions for this new subrange.

Since this randomized algorithm for the \textsc{Minkowski Sum
Selection} problem with two constraints starts with $N$ feasible
points, after the first successful contraction, we have a new range
$[s_{l'},s_{r'}]$ contains $O(N/\sqrt{n})$ feasible points and
$s^*$, the point of the $k'$-th largest $y$-coordinate among all
feasible points. After the second successful contraction, we have an
another subrange $[s_{l''},s_{r''}]$ which contains
$O(\frac{N/\sqrt{n}}{\sqrt{n}})=O(n)$ feasible points and $s^*$, the
point of the $k''$-th largest $y$-coordinate among all feasible
points. Since the number of feasible points contained in the range
$[s_{l''},s_{r''}]$ is $O(n)$, we can enumerate all feasible points
in this range in $O(n\log n)$ time via the subroutine for the
reporting version of the \textsc{Minkowski Sum Range Query} problem
with two constraints and select the point of the $k''$-th largest
$y$-coordinate from these feasible points by using any standard
selection algorithm in $O(n)$ time.

Now we show that with high probability, the point of the $k^{th}$
largest $y$-coordinate among all feasible points contained in
subrange $[s_{l'},s_{r'}]$ and the subrange $[s_{l'},s_{r'}]$
contains at most $t^2N/(t-1)\sqrt{n}$ feasible points. Applying the
results of Matou$\check{\mbox{s}}$ek~\textit{et al.}~\cite{MMN}, we
have the following lemma:


%

\begin{lemma}
Given a set of feasible points
$\Theta=\{\theta_1,\theta_2,\ldots,\theta_N\}$, an index $k$ $(1\leq
k\leq N)$, and an integer $n>0$, we can compute in $O(n\log n)$ time
an interval $[s_{l'},s_{r'}]$, such that, with probability
$1-1/\Omega(\sqrt{n})$, the point of the $k^{th}$ largest
$y$-coordinate among $\Theta$ lies within this interval, and the
number of points in $\Theta$ that lie within the interval is at most
$N/\Omega(\sqrt{n})$.
\end{lemma}

By the results given by Matou$\check{\mbox{s}}$ek~\textit{et
al.}~\cite{MMN}, we can choose $t=3$. Therefore, we just need to
repeat the contraction step at most twice on average in the
randomized algorithm for the \textsc{Minkowski Sum Selection}
problem with two constraints. We conclude the time complexity of the
randomized algorithm in the following theorem.

\begin{theorem}
For linear objective functions, the \textsc{Minkowski Sum Selection}
problem with two linear constraints can be solved in expected
$O(n\log n)$ time.
\end{theorem}

\section{Appendix E: Minkowski Sum Selection with $\lambda>2$ Constraints}
In the following, we describe how to solve the \textsc{Minkowski Sum
Selection} problem with $\lambda>2$ constraints and a linear
objective function by using Algorithm Selection$_2$ and Algorithm
Ranking$_2$. The pseudocodes of Algorithm Selection$_2$ and Algorithm
Ranking$_2$ are given in Figure~\ref{algo4} and Figure~\ref{algo5}, respectively.
The running time of Algorithm Selection$_2$ is $O(n\log^2 n)$ by Theorem~\ref{Selection2time}, and the running time of
Algorithm Ranking$_2$ is $O(n\log n)$.

    \begin{figure}[!h]
    \noindent\hrulefill\vspace{-0.6cm}
    \begin{tabbing}
    \hspace*{1em} \= \hspace*{1em} \= \hspace*{1em} \= \hspace*{1em}
    \=
    \hspace*{1em} \kill \\
    \textbf{Algorithm} {Selection$_2(P,Q,L_1,L_2,f,k)$}\\
    \textbf{Input:} $P\subseteq \mathbb{R}^2$ and $Q\subseteq \mathbb{R}^2$ are two multisets; $L_1$ and $L_2$ are
    linear\\
    \>constraints; $f:{\mathbb{R}^{2}}\rightarrow \mathbb{R}$ is a linear objective
    function; $k$ is a positive integer.\\
    \textbf{Output:} The $k^{th}$ largest value among
    all objective values of points in $(P\oplus Q)_{L_1,L_2}$.\\

    \ 1\> Perform the input transformation in Section~4.1. Let $L_1:x\geq c_1$ and\\
    \>$L_2:ax+by\geq c_2$ be the resulting constraints after the input transformation.\\
     \> /*Assume that $a<0$ and $b<0$.*/\\
    \ 2\> Sort $P$ and $Q$ into $P_y$ and $Q_y$, respectively, in nondecreasing order of $y$-coordinates.\\
    \ 3\> Let  $P_y = ((x'_1,y'_1),\ldots,(x'_{n},y'_{n}))$ and $Q_y = ((\bar{x}'_1,\bar{y}'_1),\ldots,(\bar{x}'_{n},\bar{y}'_{n}))$.\\
    \ 4\> Denote by $Y$ the sorted matrix of dimensions $n\times n$ where the $(i,j)$-th element\\
    \>is $y'_i+\bar{y}'_j$.\\
    \ 5\> $u\leftarrow n^2$; $l\leftarrow 1$; $m\leftarrow \frac{u-l+1}{2}$; $t \leftarrow$ the $m^{th}$ largest element of $Y$.\\
    \ 6\> \textbf{while} true \textbf{do}\\
    \ 7\qquad \textbf{else}\\
    \ 8\qquad\qquad $R\leftarrow$ Ranking$_2(P,Q,L_1,L_2,f,t)$.\\
    \ 9\qquad\qquad \textbf{if} $R<k$ \textbf{then}\\
     10\qquad\qquad\qquad $l \leftarrow m$; $m\leftarrow \frac{u+l+1}{2}$; $t \leftarrow$ the $m^{th}$ largest element of $Y$.\\
     11\qquad\qquad \textbf{else if} $R=k$ \textbf{then}\\
     12\qquad\qquad\qquad\quad Find the point $p=(x^*,y^*)$ in $(P\oplus Q)_{L_1,L_2,x\leq t}$ such that $y^*$\\
     \>\qquad\qquad\qquad\quad is closest to $t$.\\
     13\qquad\qquad\qquad\quad \textbf{return} $y^*$.\\
     14\qquad\qquad \textbf{else}\\
     15\qquad\qquad\qquad $u \leftarrow m$; $m\leftarrow \frac{u-l+1}{2}$; $t \leftarrow$ the $m^{th}$ largest element of $Y$.\\
    \end{tabbing}
    \vspace{-0.75 cm}\noindent\hrulefill
    \caption{The algorithm for the \textsc{Minkowski Sum Selection} problem with two linear constraints and a linear objective function.}
    \label{algo4}
    \end{figure}

\begin{figure}[!h]
    \noindent\hrulefill\vspace{-0.6cm}
    \begin{tabbing}
    \hspace*{1em} \= \hspace*{1em} \= \hspace*{1em} \= \hspace*{1em}
    \=
    \hspace*{1em} \kill \\
    \textbf{Algorithm} {Ranking$_2(P,Q,L_1,L_2,f,t)$}\\
    \textbf{Input:} Two multisets $P\subseteq \mathbb{R}^2$ and $Q\subseteq \mathbb{R}^2$; two linear constraints $L_1$ and $L_2$;\\
    \>a linear objective function $f:{\mathbb{R}^{2}}\rightarrow \mathbb{R}$; a real number $t$.\\
    \textbf{Output:} The rank of $t$ among the objective values of points in $(P\oplus
    Q)_{L_1,L_2}$.\\

    \ 1\> Perform the input transformation in Section~4.1. Let $L_1:x\geq c_1$ and\\
    \>$L_2:ax+by\geq c_2$ be the resulting constraints after the input transformation.\\
     \>/*Assume that $a<0$ and $b<0$.*/\\
     \ 2\> Sort $P$ and $Q$ into $P_y$ and $Q_y$, respectively, in nondecreasing order of
    $y$-coordinates.\\
    \ 3\> Let  $P_y = ((x'_1,y'_1),\ldots,(x'_{n},y'_{n}))$ and $Q_y = ((\bar{x}'_1,\bar{y}'_1),\ldots,(\bar{x}'_{n},\bar{y}'_{n}))$.\\
    \ 4\> Denote by $Y$ the sorted matrix of dimensions $n\times n$ where the $(i,j)$-th
    element\\
    \>is $y'_i+\bar{y}'_j$.\\
    \ 5\> $R_1 \leftarrow$ Ranking$_1(P,Q,L'$: $x<c_1,f'(x,y)=y,t)-1$.\\
    \ 6\> $R_2 \leftarrow$ Ranking$_1(P,Q,L''$: $ax+by<c_2,f'(x,y)=y,t)-1$.\\
    \ 7\> $R_3 \leftarrow$ Ranking$_1(P,Q,L''$: $ax+by<c_2,f''(x,y)=-x,c_1)-1$.\\
    \ 8\> $R_t \leftarrow$ the rank of $t$ among the values of $y$-coordinates of the points in $Y$ minus one.\\
    \ 9\> $R \leftarrow R_t-R_1-R_2+R_3+1$.\\
     10\> \textbf{return} $R$.\\
    \end{tabbing}
    \vspace{-0.75 cm}\noindent\hrulefill
    \caption{The ranking algorithm for the Minkowski sum with two linear constraints and a linear objective function.}
    \label{algo5}
    \end{figure}

\begin{figure}
\centering
\includegraphics[scale=0.45]{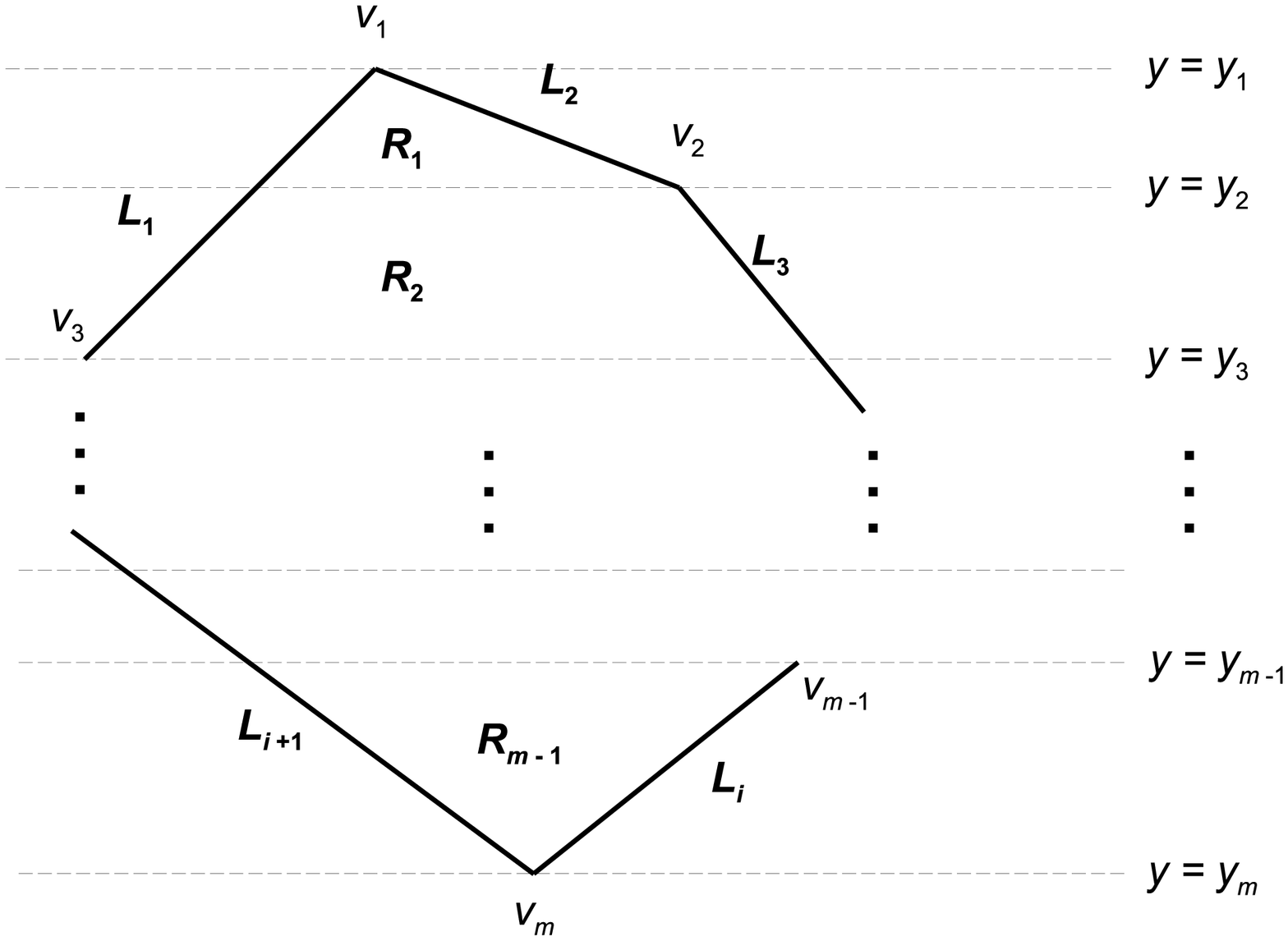}
\caption{The polygon formed by the $\lambda$
constraints.} \label{polygon}
\end{figure}

Without loss of generality, we assume the given objective function
is $f(x,y)=y$; otherwise, we may perform some transformations on the
input first. Before moving on to the algorithm, let us pause here to
introduce some definitions. Let
$\chi=\{L_1,L_2,\ldots,L_{\lambda}\}$ the set of the $\lambda$
constraints and $V=(v_1 = (x_1,y_1),v_2 =
(x_2,y_2),\ldots,v_m=(x_m,y_m))$ be the vertices of the polygon
formed by the $\lambda$ constraints, sorted in nonincreasing order
of $y$-coordinates. For simplicity, we assume that each point in
$(P\oplus Q)_\chi$ and $V$ has a distinct $y$-coordinate and the
polygon formed by the $\lambda$ constraints is closed. If not, we
can add another constraint making that the polygon is closed and
contains all points in $(P\oplus Q)_\chi$. Denote by $L_1^i$ and
$L_2^i$ the two constraints resulting in the edges of the polygon
between lines $y=y_i$ and $y=y_{i+1}$ for each $i=1,\ldots, m-1$.
For each $i =1,2,\ldots,m-1$, we define $R_i$ to be the point set
$\{(x,y):(x,y)\in (P\oplus Q)_\chi \mbox{ and } y_{i+1} < y \leq
y_i\}$. An illustration of the above definitions is shown in
Figure~\ref{polygon}.

We now explain how our algorithm works. First of all, we have to
compute the value of $|R_i|$ for each $i=1,2,\ldots,m-1$. The value
of $|R_i|$ is equal to the number of points in $(P\oplus
Q)_{L_1^i,L_2^i}$ above the line $y=y_{i+1}$ minus the number of
points in $(P\oplus Q)_{L_1^i,L_2^i}$ above the line $y=y_i$. The
number of points in $(P\oplus Q)_{L_1^i,L_2^i}$ above the line
$y=y_{i+1}$ is equal to the rank of $y_{i+1}$ among the values of
$y$-coordinates of the points in $(P\oplus Q)_{L_1^i,L_2^i}$ minus
one, i.e., Ranking$_2(P,Q,L_1^i,L_2^i,f(x,y)=y,y_{i+1})-1$. The
number of points in $(P\oplus Q)_{L_1^i,L_2^i}$ above the line
$y=y_i$ is equal to the rank of $y_{i}$ among the values of
$y$-coordinates of the points in $(P\oplus Q)_{L_1^i,L_2^i}$ minus
one, i.e., $\mbox{Ranking}_2(P,Q,L_1^i,L_2^i,f(x,y)=y,y_i)-1$. Thus,
we have
$|R_i|=\mbox{Ranking}_2(P,Q,L_1^i,L_2^i,f(x,y)=y,y_{i+1})-\mbox{Ranking}_2(P,Q,L_1^i,L_2^i,f(x,y)=y,y_i)$
for each $i=1,2,\ldots,m-1$. After computing all values of $|R_i|$
for each $i=1,2,\ldots,m-1$, we can determine in which set $R_w$ the
solution $y^*$ is located.

Next, we explain how to determine this set $R_w$ where the solution
$y^*$ is located. Let $S_0=0$ and $S_i=|R_1|+|R_2|+\cdots+|R_{i}|$
for each $i=1,2,\ldots,m-1$. We find the smallest index $w$ such
that $S_w\geq k$ or $w=m-1$. It is easy to see that the solution
$y^*$ must be located in the set $R_w$. It follows that the solution
$y^*$ must be among the $y$-coordinates of points in $(P\oplus
Q)_{L_1^w,L_2^w}$.

The last step is to obtain the rank of $y^*$ among the
$y$-coordinates of points in $(P\oplus Q)_{L_1^w,L_2^w}$ and then we
can use Algorithm Selection$_2$ to find the solution $y^*$. Because
the solution $y^*$ is located in the set $R_w$, the rank of $y^*$
among the $y$-coordinates of points in $R_w$ is $k-S_{w-1}$. We call
$\mbox{Ranking}_2(P,Q,L_1^w,L_2^w,f(x,y)=y,y_w)$ and let $r$ be the
return value of the invocation. The value $r-1$ is equal to the
number of points in $(P\oplus Q)_{L_1^w,L_2^w}$ that are above the
line $y=y_w$. Hence, there are $r$ points in $(P\oplus
Q)_{L_1^w,L_2^w}$ that are above the line $y=y_w$ and there are
$k-S_{w-1}$ points in $(P\oplus Q)_{L_1^w,L_2^w}$ belonging to the
set $R_w$. It is easy to derive that the value $r-1+k-S_{w-1}$ is
the rank of the solution $y^*$ among the $y$-coordinates of points
in $(P\oplus Q)_{L_1^w,L_2^w}$. Finally, we call
Selection$_2(P,Q,L_1^w,L_2^w,f(x,y)=y,r-1+k-S_{w-1})$ to find the
solution $y^*$. The detailed algorithm is given in
Figure~\ref{algo6}.

\begin{figure}[!h]
    \noindent\hrulefill\vspace{-0.6cm}
    \begin{tabbing}
    \hspace*{1em} \= \hspace*{1em} \= \hspace*{1em} \= \hspace*{1em}
    \=
    \hspace*{1em} \kill \\
    \textbf{Algorithm} {Selection$_{\lambda}(P,Q,\chi,k)$}\\
    \textbf{Input:} Two multisets $P\subseteq \mathbb{R}^2$, $Q\subseteq \mathbb{R}^2$; a set $\chi$ of the $\lambda$ constraints; a positive integer $k$.\\
    \textbf{Output:} The $k^{th}$ largest value of $y$-coordinates among
    all points in $(P\oplus Q)_{L_1,\ldots,L_{\lambda}}$.\\
    \ 1\> $V\leftarrow$ vertices of the polygon formed by the $\lambda$ constraints in $\chi$.\\
    \ 2\> $m\leftarrow |V|$; $S_0\leftarrow 0$.\\
    \ 3\> \textbf{for} $i\leftarrow 1$ to $m-1$ \textbf{do}\\
    \ 4\qquad $|R_i|\leftarrow$ Ranking$_2(P,Q,L_1^i,L_2^i,f(x,y)=y,y_{i+1})$\\
    \qquad\qquad\qquad $-\mbox{Ranking}_2(P,Q,L_1^i,L_2^i,f(x,y)=y,y_i)$.\\
    \ 5\qquad $S_i\leftarrow |R_i|+S_{i-1}$.\\
    \ 6\> Find the smallest index $w$ such that $S_w\geq k$ or $w=m-1$.\\
    \ 7\> $r\leftarrow \mbox{Ranking}_2(P,Q,L_1^w,L_2^w,f(x,y)=y,y_w)$.\\
    \ 8\> \textbf{return} Selection$_2(P,Q,L_1^w,L_2^w,f(x,y)=y,r-1+k-S_{w-1})$.\\
    \end{tabbing}
    \vspace{-0.75 cm}\noindent\hrulefill
    \caption{The algorithm for the \textsc{Minkowski Sum Selection} problem with $\lambda>2$ linear constraints an a linear objective function.}
    \label{algo6}
    \end{figure}

Now let us see the running time. Because there are $\lambda$ constraints, it takes
$O(\lambda\log\lambda)$ time to compute $V$~\cite{Berg}. To compute the values of $|R_i|$ for all $i\in\{1,2,\ldots,m-1\}$, we call Algorithm Ranking$_2$ $2\lambda$ times, which takes
$O(\lambda\cdot n\log n)$ time. The invocation of Algorithm Selection$_2$ takes
$O(n\log^2 n)$ time by Theorem~\ref{Selection2time}. Hence, the
total time complexity is $O(\lambda \cdot n \log n
+\lambda \log \lambda+n\log^2 n)$. Note that if we use the
randomized algorithm described in Theorem~\ref{randomized_time}
instead of Algorithm Selection$_2$, the total time complexity is $O(\lambda
\cdot n \log n +\lambda \log \lambda+n\log n)$.
Hence, we have the following theorem.

\begin{theorem}
Let $\lambda$ be any fixed integer larger than two. The
\textsc{Minkowski Sum Selection} problem with $\lambda$ constraints
and a linear objective function is asymptotically equivalent to the
\textsc{Minkowski Sum Selection} problem with two linear constraints
and a linear objective function.
\end{theorem}


\begin{thebibliography}{99}

    \bibitem{All}
    Allison, L.: Longest Biased Interval and Longest Non-negative Sum Interval.
    Bioinformatics Application Note 19(10), 1294--1295 (2003)

    \bibitem{Ben-Or}
    Ben-Or, M.: Lower Bounds for Algebraic Computation Trees. In: Proc. STOC, pp. 80--86 (1983)

    \bibitem{Bengtsson}
    Bengtsson, F. and Chen, J.: Efficient Algorithms for $k$ Maximum Sums.
    Algorithmica 46(1), 27--41 (2006)

    \bibitem{Berg}
    Berg, M., Kreveld, M., Overmars, M., Rivest, R.L., and Schwarzkopf
    O.: Computational Geometry: Algorithms and Applications.
    Springer (2000)

    \bibitem{Bern}
    Bernholt, T., Eisenbrand, F., and Hofmeister, T.:
    A Geometric Framework for Solving Subsequence Problems in Computational Biology Efficiently.
    In SoCG, pp. 310--318 (2007)

    \bibitem{starmap}
    Chen, K.-Y. and Chao, K.-M.:
    Optimal Algorithms for Locating the Longest and Shortest Segments Satisfying a Sum or an Average Constraint.
    Information Processing Letter 96(6), 197--201 (2005)




    \bibitem{Cole}
    Cole, R., Salowe, J.S., Steiger, W.L., and Szemeredi, E.: An Optimal-Time Algorithm for Slope Selection. SIAM Journal on Computing 18(4), 792--810
    (1989)


    \bibitem{Frede}
    Frederickson, G.N. and Johnson, D.B.: Generalized Selection and Ranking: Sorted Matrices.
    SIAM Journal on Computing 13(1), 14--30 (1984)

    \bibitem{Gold}
    Goldwasser, M.H., Kao, M.-Y., and Lu, H.-I:
    Linear-time Algorithms for Computing Maximum-density Sequence Segments with Bioinformatics Applications.
    Journal of Computer and System Sciences 70(2), 128--144 (2005)

    \bibitem{Huang}
    Huang, X.: An Algorithm for Identifying Regions of a DNA
    Sequence that Satisfy a Content Requirement.
    Computer Applications in the Biosciences 10(3), 219-225 (1994)

    \bibitem{Ioshikhes}
    Ioshikhes, I. and Zhang, M.Q.: Large-Scale Human Promoter Mapping Using CpG Islands. Nature Genetics 26(1), 61-63
    (2000)


    \bibitem{LLL}
    Lee, D.T., Lin, T.-C., and Lu, H.-I: Fast Algorithms for the Density Finding Problem.
    Algorithmica DOI:10.1007/s00453-007-9023-8 (2007)

    \bibitem{TC06}
    Lin, T.-C. and Lee, D.T.: Efficient Algorithm for the Sum Selection Problem and
    $k$ Maximum Sums Problem. In: Asano T. (eds) ISAAC 2006. LNCS, vol 4288, pp. 460--473. Springer, Heidelberg
    (2006)

    \bibitem{TC05}
    Lin, T.-C. and Lee, D.T.: Randomized Algorithm for the Sum Selection Problem.
    Theoretical Computer Science 377(1-3), 151--156 (2007)

    \bibitem{LJC}
    Lin, Y.-L., Jiang, T., and Chao, K.-M.:
    Efficient Algorithms for Locating the Length-constrained Heaviest Segments with Applications to Biomolecular Sequence Analysis.
    Journal of Computer and System Sciences 65(3), 570--586 (2002)

    \bibitem{LHJC}
    Lin, Y.-L., Huang, X., Jiang, T., and Chao, K.-M.:
    MAVG: Locating Non-overlapping Maximum Average Segments in a Given Sequence. Bioinformatics 19(1), 151--152
    (2003)

    \bibitem{Lip}
    Lipson, D., Aumann, Y., Ben-Dor, A., Linial, N., and Yakhini,
    Z.:
    Efficient Calculation of Interval Scores for DNA Copy Number Data Analysis.
    Journal of Computational Biology 13(2), 215-228 (2006)

    \bibitem{Ma}
    Matou$\check{\mbox{s}}$ek, J.:
    Randomized optimal algorithm for slope selection.
    Information Processing Letters 39(4), 183--187 (1991)

    \bibitem{MMN}
    Matou$\check{\mbox{s}}$ek, J., Mount, D.M., and Netanyahu, N.S.:
    Efficient Randomized Algorithms for the Repeated Median Line Estimator.
    Algorithmica 20(2), 136--150 (1998)

    \bibitem{Ohler}
    Ohler, U., Niemann, H., Liao, G.-C., and Rubin, G.M.: Joint Modeling of DNA Sequence and Physical Properties to Improve Eukaryotic Promoter Recognition. Bioinformatics 199--206
    (2001)

    \bibitem {Wang}
    Wang, L. and Xu, Y.: SEGID: Identifying Interesting Segments in (Multiple)
    Sequence Alignments. Bioinformatics 19(2), 297-298
    (2003)

\end{thebibliography}
\end{document}